\documentclass[journal]{IEEEtran}

\usepackage{amsfonts}
\usepackage{amssymb}
\usepackage{amsthm}
\usepackage{amsmath,amsfonts,amssymb}
\usepackage[dvips]{graphicx}
\usepackage{subfigure}
\usepackage{verbatim}
\usepackage{setspace}
\usepackage{bm}
\usepackage{algorithmic} 
\usepackage[ruled,vlined]{algorithm2e}
\usepackage{cite}

\usepackage{changepage}
\usepackage{pdfpages}
\usepackage{color}

\newcommand{\figwidth}{8}
\IEEEoverridecommandlockouts

\begin{document}
\title{Movable-Antenna Enhanced Multiuser Communication via Antenna Position Optimization}
\author{Lipeng Zhu, ~\IEEEmembership{Member,~IEEE,}
		Wenyan Ma,~\IEEEmembership{Graduate Student Member,~IEEE,}
		Boyu Ning,~\IEEEmembership{Member,~IEEE,}
		and Rui Zhang,~\IEEEmembership{Fellow,~IEEE}
	\thanks{This work is supported in part by Ministry of Education, Singapore under Award T2EP50120-0024, Advanced Research and Technology Innovation Centre (ARTIC) of National University of Singapore under Research Grant R-261-518-005-720, and The Guangdong Provincial Key Laboratory of Big Data Computing. (\textit{Corresponding author: Lipeng Zhu and Rui Zhang})}
	\thanks{L. Zhu and W. Ma are with the Department of Electrical and Computer Engineering, National University of Singapore, Singapore 117583 (e-mail: zhulp@nus.edu.sg, wenyan@u.nus.edu).}
	\thanks{Boyu Ning is with National Key Laboratory of Science and Technology on Communications, University of Electronic Science and Technology of China (UESTC), Chengdu 611731, China (e-mail: boydning@outlook.com).}
	\thanks{R. Zhang is with School of Science and Engineering, Shenzhen Research Institute of Big Data, The Chinese University of Hong Kong, Shenzhen, Guangdong 518172, China (e-mail: rzhang@cuhk.edu.cn). He is also with the Department of Electrical and Computer Engineering, National University of Singapore, Singapore 117583 (e-mail: elezhang@nus.edu.sg). }
}

\maketitle

\vspace{-0.8 cm}
 

\begin{abstract}
	Movable antenna (MA) is a promising technology to improve wireless communication performance by varying the antenna position in a given finite area at the transceivers to create more favorable channel conditions. In this paper, we investigate the MA-enhanced multiple-access channel (MAC) for the uplink transmission from multiple users each equipped with a single MA to a base station (BS) with a fixed-position antenna (FPA) array. A field-response based channel model is used to characterize the multi-path channel between the antenna array of the BS and each user's MA with a flexible position. To evaluate the MAC performance gain provided by MAs, we formulate an optimization problem for minimizing the total transmit power of users, subject to a minimum-achievable-rate requirement for each user, where the positions of MAs and the transmit powers of users, as well as the receive combining matrix of the BS are jointly optimized. To solve this non-convex optimization problem involving intricately coupled variables, we develop two algorithms based on zero-forcing (ZF) and minimum mean square error (MMSE) combining methods, respectively. Specifically, for each algorithm, the combining matrix of the BS and the total transmit power of users are expressed as a function of the MAs' position vectors, which are then optimized by using the proposed multi-directional descent (MDD) framework. It is shown that the proposed ZF-based and MMSE-based MDD algorithms can converge to high-quality suboptimal solutions with low computational complexities. Simulation results demonstrate that the proposed solutions for MA-enhanced multiple access systems can significantly decrease the total transmit power of users as compared to conventional FPA systems employing antenna selection under both perfect and imperfect field-response information. 
\end{abstract}

%
\IEEEpeerreviewmaketitle

\section{Introduction}
\IEEEPARstart{W}{ith} advances in the multiple-input multiple-output (MIMO) technologies, the capacity of today's wireless communication systems has been dramatically increased by exploiting the new degrees of freedom (DoFs) in the spatial domain. By leveraging the independent/quasi-independent channel fading caused by the random superposition of multipath components, MIMO systems can support parallel transmissions of multiple data streams in the same time-frequency resource block \cite{telatar1999capacity,Paulraj2004Anover,Stuber2004broadb}. Thus, MIMO and/or massive MIMO technologies can improve the spectral efficiency manifold compared to single-antenna systems. However, conventional MIMO and/or massive MIMO systems usually adopt fixed-position antennas (FPAs) with spacing no smaller than a half-wavelength at the transceivers \cite{Larsson2014mMIMO,Lu2014Anover,zeng2016millim,zhu2019millim,Ning2023THzbeam}. Such fixed and discrete deployment of antennas limits the diversity and spatial multiplexing performance of MIMO systems because the channel variation in the continuous spatial field is not fully utilized.

Recently, movable antenna (MA) was proposed as a new solution to break the above fundamental limitations of FPA systems \cite{zhu2022MAmodel, ma2022MAmimo}. Specifically, an MA is connected to the radio frequency (RF) chain via a flexible cable and it can be moved in a spatial region with the aid of a driver. The implementation of an MA system is similar to the widely explored distributed antenna systems \cite{Castanheira2010distri,Heath2013Acurrent,Cui2019greend}, but the former employs much shorter connecting cables as the size of spatial region for antenna moving is in the order of several to tens of wavelengths only \cite{Zhuravlev2015experi,Basbug2017design}. For example, the authors in \cite{Zhuravlev2015experi} presented a motor-enabled MA for realizing multi-static radar, where the transmit and receive antennas can be moved along parallel lines independently with the aid of mechanical scanners. In \cite{Basbug2017design}, a reconfigurable antenna array was designed with its array elements moving along a semicircular path to synthesis radiation patterns. The authors in \cite{Li2022movingAnte} proposed to integrate an MA on a ground robotic vehicle for improving the observability of the inertial navigation system (INS) and global navigation satellite system (GNSS). In \cite{eliyahu2022single}, a proof-of-concept hardware was designed for direction finding by using a single moving omnidirectional antenna. In \cite{Do2021reconf}, a reconfigurable uniform linear array (ULA) was proposed to approach the line-of-sight (LoS) MIMO capacity by enabling the rotation of the antenna array, which can be regarded as a practical way of implementing MAs in a restricted area. 

For wireless communication systems, an MA can be fast deployed to the position with more favorable channel conditions for improving the communication performance between the transmitter (Tx) and receiver (Rx). Compared to conventional FPA systems, the MA-enabled communication systems have some appealing advantages \cite{zhu2022MAmodel, ma2022MAmimo,zhu2023MAMag}. First, the MA systems can reap the full diversity in the given spatial region. Different from conventional FPAs which undergo random and uncontrolled channel fading, the MAs at Tx and/or Rx can be deployed at positions that achieve the highest channel gain between Tx and Rx. As such, the small-scale fading of the channel in the spatial domain is fully exploited to increase the signal-to-noise ratio (SNR) at the Rx. Second, the MA systems can provide interference mitigation gain. Due to the capability of flexible movement, the MA at Rx can be deployed at the position which experiences a deep fading channel with the interference source. Thus, the signal-to-interference-plus-noise ratio (SINR) at Rx can be significantly increased by leveraging the spatial DoFs even without multiple antennas. Third, the MA-enabled MIMO systems can achieve higher spatial multiplexing rate. By optimizing the positions of multiple MAs, the channel matrix between Tx and Rx can be reshaped such that the MIMO capacity is maximized.

Preliminary studies on MA-enabled communication systems have validated their performance gain over conventional FPA systems in various system setups \cite{zhu2022MAmodel,ma2022MAmimo,zhu2023MAMag,zhu2023MAarray,Wong2021fluid,Wong2022fluid,chai2022protsel,khammassi2022newchann}. In \cite{zhu2022MAmodel}, the mechanical MA architecture and a field-response based channel model for single-MA systems were proposed, where the conditions under which the field-response based channel model becomes the well-known LoS channel, geometric channel, Rayleigh and Rician fading channel models were discussed. In addition, the SNR gain achieved by a single receive MA over its FPA counterpart was analyzed under both deterministic and stochastic channels. The analytical and simulation results revealed that the performance gain of MA systems is highly depended on the number of channel paths and the size of the spatial region for moving the antenna. In \cite{ma2022MAmimo}, the MA-enabled MIMO system was investigated, where the positions of MAs at Tx and Rx were jointly optimized with the covariance matrix of transmit signals for maximizing the channel capacity. To address this non-convex problem, an alternating optimization algorithm was developed by iteratively optimizing the position of each transmit/receive MA and the transmit covariance matrix with the other variables being fixed. Simulation results showed that the MA-enabled MIMO system can significantly increase the channel capacity compared to conventional FPA-enabled MIMO systems with/without antenna selection. In \cite{zhu2023MAMag}, the motor-based architecture for implementing MAs was presented, where the MA is installed on a three-dimensional (3D) mechanical slide driven by step motors. In addition, an overview of the application scenarios, technical potentials, challenges and solutions for MA-aided wireless communications was given in \cite{zhu2023MAMag}. The authors in \cite{Wong2021fluid} proposed an alternative implementation of MA, namely fluid antenna system (FAS), where the antenna is fabricated by using liquid metals or ionized solutions \cite{paracha2019liquid,dong2021liquid,morishita2013liquid}, and it can be placed at one of the candidate ports in a one-dimensional (1D) line space. By assuming uniform scattering environments, the outage probability of the single-antenna FAS was derived based on an approximation of the spatially-correlated Rayleigh fading channels. Then, the results were extended to multiuser systems in \cite{Wong2022fluid}, where multiple transceivers can send/receive signals simultaneously with the interference mitigated by switching the physical location of the fluid antenna at each user's Rx. It was shown that by using one fluid antenna moving in a line space of a few wavelengths, hundreds of users can communicate at the same time if the channel fading is sufficiently prominent in the spatial domain. In \cite{chai2022protsel}, the port selection for FAS was studied for approaching the maximum SNR at Rx, where the machine learning method was used to capture the implicit channel correlation between closely-spaced antenna ports so as to reduce the number of port observations. Considering the inaccurate channel model adopted in \cite{Wong2021fluid,Wong2022fluid,chai2022protsel}, the authors in \cite{khammassi2022newchann} proposed a new analytical approximation of the FAS channel, and the FAS channel model was shown to achieve a good performance on approximating the spatial correlation between the antenna ports, i.e., Jake's model.

The above two types of implementation for MA, i.e., the mechanical MA proposed in \cite{zhu2022MAmodel,ma2022MAmimo} and the FAS shown in \cite{Wong2021fluid,Wong2022fluid,chai2022protsel,khammassi2022newchann}, both exploit the diversity gain and the interference mitigation gain in the spatial domain by adjusting the positions of MAs. However, they differ significantly in terms of hardware architecture and channel model. On one hand, the mechanical MA adopted in  \cite{zhu2022MAmodel,ma2022MAmimo} can achieve a more flexible movement in the two-dimensional (2D) or 3D space, but it requires additional hardware cost for installing drivers to move antennas. In comparison, the FAS is easy to be integrated into a small area, but due to the liquid form, an antenna can only be moved in a 1D line space and it is difficult to form antenna arrays. On the other hand, in \cite{Wong2021fluid,Wong2022fluid,chai2022protsel,khammassi2022newchann}, the port-based channel model in a 1D line space under the uniform scattering assumption may not be applicable to practical systems with finite number of scatters and multi-paths in the 3D space. Due to the discrete form of the port-based channel model, the works in \cite{Wong2021fluid,Wong2022fluid,chai2022protsel,khammassi2022newchann} can only exhaustively search the optimal antenna ports/positions in a discrete set, which is similar to the methodology adopted in conventional antenna selection systems. In contrast, the field-response based channel model proposed in \cite{zhu2022MAmodel} characterizes the variation of the channel between Tx and Rx in a continuous 1D/2D/3D spatial region under the far-field condition. Under such channel models, the antenna position can be optimized in continuous Tx/Rx regions, which provides a new paradigm for designing wireless communication systems.

In light of the above, this paper considers the mechanical MA structure and field-response based channel model to investigate the multiple-access channel (MAC) enhanced by MAs. Specifically, multiple users each equipped with a single MA are served by a based station (BS) equipped with an FPA array. Although the multi-antenna multiuser communication systems have been widely investigated in existing literature \cite{Vishwanath2003duality,Yu2006duality,Shi2011WMMSE,Sohrabi2016Hybrid,Ho2019multiuser}, these works mainly focused on FPA systems, while the DoF in antenna position optimization was not considered or fully exploited. To evaluate the performance gain provided by MAs, it is essential to characterize the channel variation between the antenna array of the BS and each user's antenna moving in a given 3D spatial region. Thus, the field-response based channel model in \cite{zhu2022MAmodel} is adopted to enable the MA position optimization for multiple users, whereas the channel model is extended from 2D surface to 3D space. Based on this channel model, we formulate an optimization problem for minimizing the total transmit power of users, subject to a minimum-achievable-rate requirement for each user in the uplink, where the positions of MAs and the transmit power of users, as well as the receive combining matrix of the BS are jointly optimized. Since the resultant problem is non-convex and involves highly coupled variables, we develop two algorithms based on zero-forcing (ZF) and minimum mean square error (MMSE) combining methods to obtain high-quality suboptimal solutions with low computational complexities. For each algorithm, the combining matrix of the BS and the transmit power of users are expressed as a function of MAs' position vectors, which are then optimized by using the proposed multi-directional descent (MDD) framework. The convergence of the proposed two algorithms is analyzed, and an alternative solution for the single-user case is also presented. Simulation results validate the efficacy of the proposed MA-enhanced multiple access systems, where the total transmit power of multiple users can be significantly decreased compared to conventional FPA systems employing antenna selection (AS). The results also reveal that the interference mitigation gain of MA systems over FPA systems becomes more pronounced for larger number of users and higher achievable-rate targets. Besides, the performance gain of MAs is highly depended on the number of channel paths between the BS and users as well as the size of the spatial region for moving antennas. Moreover, we evaluate the impact of imperfect field-response information (FRI) on the solution for MA positioning. It is shown that the proposed algorithms can achieve a robust performance even if the estimated angles and coefficients of the channel paths are inaccurate.

The rest of this paper is organized as follows. Section II introduces the signal model and the field-response based channel model for the MA-enhanced multiple access system, and then presents the problem formulation for MAs' position optimization. In Section III, we show the ZF-based and MMSE-based MDD solutions for solving the optimization problem, where the convergence and computational complexities are analyzed. Simulation results and main observations are provided in Section IV and finally this paper is concluded in Section V.

\textit{Notation}: $a$, $\mathbf{a}$, $\mathbf{A}$, and $\mathcal{A}$ denote a scalar, a vector, a matrix, and a set, respectively. $\mathbf{A}^{\rm{T}}$ and $\mathbf{A}^{\rm{H}}$ denote the transpose and the conjugate transpose of matrix $\mathbf{A}$, respectively. $\mathbf{A}^{-1}$ and $\mathbf{A}^{\dagger}$ are the inverse and the pseudo-inverse of matrix $\mathbf{A}$, respectively. $\|\mathbf{a}\|_{1}$, $\|\mathbf{a}\|_{2}$ and $\|\mathbf{A}\|_{\mathrm{F}}$ represent the 1-norm of vector $\mathbf{a}$, the 2-norm of vector $\mathbf{a}$, and the Frobenius norm of matrix $\mathbf{A}$, respectively. $[\mathbf{a}]_i$, $[\mathbf{A}]_{i,j}$, and $[\mathbf{A}]_{:,j}$ denote the $i$-th entry of vector $\mathbf{a}$ and the entry in the $i$-th row and $j$-th column of matrix $\mathbf{A}$, and the $j$-th column vector of matrix $\mathbf{A}$, respectively. $\operatorname{diag}\{\mathbf{a}\}$ is a diagonal matrix with the entry in the $i$-th row and $i$-th column equal to the $i$-th entry of vector $\mathbf{a}$. $\operatorname{tr}\{\mathbf{A}\}$ and $\rho\{\mathbf{A}\}$ denote the trace and the spectral radius of matrix $\mathbf{A}$, respectively. $\lambda_{k}\{\mathbf{A}\}$ denotes the $k$-th eigenvalue of matrix $\mathbf{A}$. $\mathbf{0}_{N}$ denotes an $N$-dimensional vector with all elements equal to 0. $\mathbf{e}_{N}^{i}$ is an $N$-dimensional vector with the value of one for the $i$-th element and zero elements elsewhere. $\mathbf{I}_{N}$ denotes an identical matrix of size $N \times N$. $\mathcal{CN}(\mathbf{0},\mathbf{\Lambda})$ represents the circularly symmetric complex Gaussian (CSCG) distribution with mean zero and covariance matrix $\mathbf{\Lambda}$. $\mathcal{U}[a,b]$ denotes the uniform distribution within real-number interval $[a,b]$. $\mathbb{E}\{\cdot\}$ denotes the expected value of a random variable. $\mathbb{R}$ and $\mathbb{C}$ represent the sets of real and complex numbers, respectively. Operator $\partial(\cdot)$ denotes the partial differential. $\nabla_{\mathbf{x}} f(\mathbf{x})$ denotes the gradient of function $f(\mathbf{x})$ with respect to (w.r.t.) $\mathbf{x}$. 

\section{System Model and Problem Formulation}
\begin{figure*}[t]
	\begin{center}
		\includegraphics[width=12 cm]{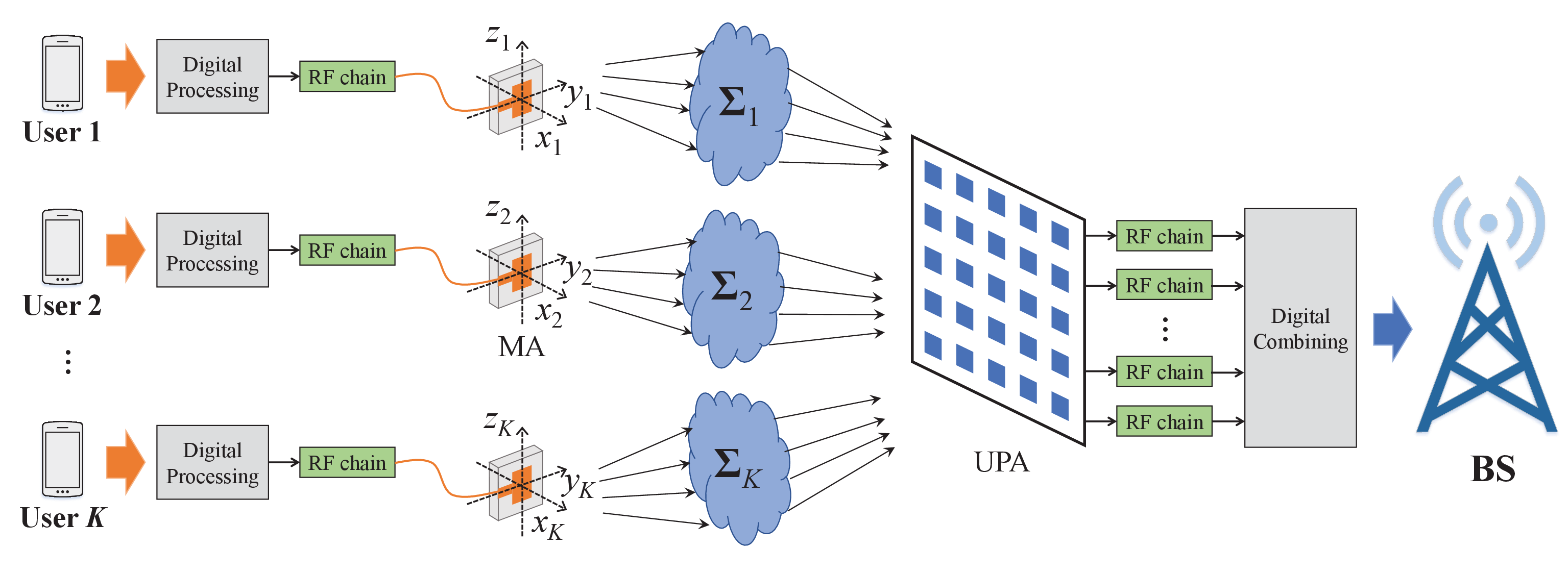}
		\caption{Illustration of the uplink transmission between $K$ single-MA users and the BS equipped with a UPA of size $N=N_{1} \times N_{2}$.}
		\label{fig:system}
	\end{center}
\end{figure*}

\begin{figure}[t]
	\begin{center}
		\includegraphics[width=7.0 cm]{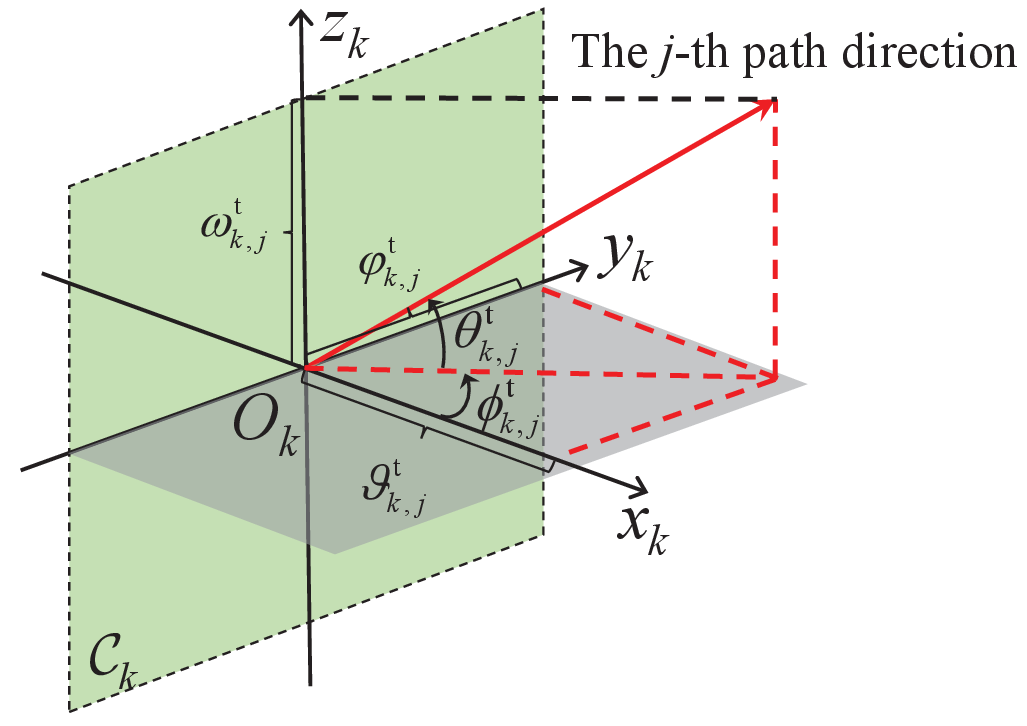}
		\caption{Illustration of the 3D local coordinate system and the corresponding spatial angles for user $k$, $1 \leq k \leq K$.}
		\label{fig:Coordinates}
	\end{center}
\end{figure}

As shown in Fig. \ref{fig:system}, the BS is equipped with a uniform planar array (UPA) of size $N=N_{1} \times N_{2}$ to serve $K$ single-MA users, where $N_{1}$ and $N_{2}$ denote the number of antennas along horizontal and vertical directions, respectively. We assume that the number of users does not exceed that of antennas at the BS, i.e., $K \leq N$, and thus the space-division multiple access (SDMA) can be used. For each user $k$, the MA is connected to the RF chain via a flexible cable such that it can be moved in a local region $\mathcal{C}_{k}$. A 3D local coordinate system shown in Fig. \ref{fig:Coordinates} is established to describe the position of MA for user $k$, which is denoted as $\mathbf{u}_{k}=[x_{k},y_{k},z_{k}]^{\mathrm{T}} \in \mathcal{C}_{k}$, $1 \leq k \leq K$. Without loss of generality, we assume that the local region for moving the antenna is a cuboid, i.e., $\mathcal{C}_k=\left[x_k^{\min }, x_k^{\max }\right] \times\left[y_k^{\min }, y_k^{\max }\right] \times\left[z_k^{\min }, z_k^{\max }\right]$, $1 \leq k \leq K$\footnote{In reality, the positioning accuracy of MAs can be up to hundredth/thousandth of the signal wavelength for existing wireless communication systems, e.g., by installing MAs on mechanical slides driven by step motors \cite{zhu2023MAMag}. As such, the channel variation due to the MA positioning error can be neglected and the assumption of MA continuous movement is adopted in this paper.}. Note that the local coordinate systems for multiple users are independently defined, with the origin of the $k$-th user being $O_{k}$. Besides, the local coordinate of the $n$-th FPA at the BS is denoted as $\mathbf{v}_{n}=[X_{n},Y_{n},Z_{n}]$, $1 \leq n \leq N$.

Let $\mathbf{h}_{k}(\mathbf{u}_{k}) \in \mathbb{C}^{N \times 1}$ denote the channel vector between the BS and user $k$, which is determined by the propagation environment and the position of the MA, $\mathbf{u}_{k}$. We consider the uplink transmission from the users to BS, where the received signals via the MAC can be expressed as\footnote{According to the uplink-downlink duality, the achievable rate region for the MAC is the same as its dual broadcast channel (BC) if the total transmit power is identical \cite{Boche2002duality,Vishwanath2003duality,Yu2006duality}. Thus, the solutions for MA positioning in the uplink are also applicable for that in the downlink.}
\begin{equation}\label{eq_signal}
	\mathbf{y} = \mathbf{W}^{\mathrm{H}}\mathbf{H}(\tilde{\mathbf{u}})\mathbf{P}^{1/2}\mathbf{s}+\mathbf{W}^{\mathrm{H}}\mathbf{n},
\end{equation}
where $\mathbf{W}=\left[\mathbf{w}_{1},\mathbf{w}_{2},\cdots,\mathbf{w}_{K}\right] \in \mathbb{C}^{N \times K}$ is the receive combining matrix of the BS, and $\mathbf{w}_{k}$ is the combining vector for user $k$, $1 \leq k \leq K$. $\mathbf{H}(\tilde{\mathbf{u}})=\left[\mathbf{h}_{1}(\mathbf{u}_{1}),\mathbf{h}_{2}(\mathbf{u}_{2}),\cdots,\mathbf{h}_{K}(\mathbf{u}_{K})\right] \in \mathbb{C}^{N \times K}$ denotes the MAC matrix from all $K$ users to the antenna array of the BS, where $\tilde{\mathbf{u}}=\left[\mathbf{u}_{1}^{\mathrm{T}},\mathbf{u}_{2}^{\mathrm{T}},\cdots,\mathbf{u}_{K}^{\mathrm{T}}\right]^{\mathrm{T}} \in \mathbb{R}^{3K \times 1}$ is the MA positioning vector for the $K$ users. $\mathbf{s}=\left[s_{1},s_{2},\cdots,s_{K}\right]^{\mathrm{T}}\in \mathbb{C}^{K \times 1}$ represents the transmitted signals of the users with normalized power, i.e., $\mathbb{E}\{\mathbf{s}^{\mathrm{H}}\mathbf{s}\}=\mathbf{I}_{K}$. $\mathbf{P}^{1/2}=\operatorname{diag}\left\{\sqrt{p}_{1},\sqrt{p}_{2},\cdots,\sqrt{p}_{K}\right\}$ denotes the power scaling matrix with the transmit power of user $k$ given by $p_{k}$, $1 \leq k \leq K$. $\mathbf{n}=\left[n_{1},n_{2},\cdots,n_{N}\right]^{\mathrm{T}} \sim \mathcal{CN}(\mathbf{0},\sigma^{2}\mathbf{I}_{N})$ is the additive white Gaussian noise (AWGN) at the BS with $\sigma^{2}$ being the average noise power.

\subsection{Channel Model}
In this paper, we consider narrow-band channels with slow fading and focus on one quasi-static fading block. As we have mentioned, the channel vector between each user and the BS is determined by the propagation environment and the position of MA. For simplicity, we assume that the far-field condition is satisfied between the BS and users, where the sizes of moving regions ($\mathcal{C}_{k}$) for the users and the UPA of the BS are much smaller than the signal propagation distance. Under this assumption, the plane-wave model can be used to form the field response from the MA region of each user to the UPA of the BS. In particular, the angle of departure (AoD), the angle of arrival (AoA), and the amplitude of the complex coefficient for each channel path between the BS and each user do not change for different positions of the MA in its corresponding region\footnote{In this paper, we assume that each MA is omnidirectional and has a fixed orientation at each user, where only the position of the MA can be changed. The joint optimization of MA's position and orientation will be an interesting topic worthy of further investigation.}, while only the phase of the channel path varies with the MA position \cite{zhu2022MAmodel}. 

Let $L_{k}^{\mathrm{t}}$ and $L_{k}^{\mathrm{r}}$, $1 \leq k \leq K$, denote the total number of transmit and receive channel paths from user $k$ to the BS, respectively. As shown in Fig. \ref{fig:Coordinates}, the elevation and azimuth AoDs for the $j$-th transmit path between user $k$ and the BS are denoted as $\theta_{k,j}^{\mathrm{t}}$ and $\phi_{k,j}^{\mathrm{t}}$, $1 \leq j \leq L_{k}^{\mathrm{t}}$, respectively. The elevation and azimuth AoAs for the $i$-th receive path between user $k$ and the BS are denoted as $\theta_{k,i}^{\mathrm{r}}$ and $\phi_{k,i}^{\mathrm{r}}$, $1 \leq i \leq L_{k}^{\mathrm{r}}$, respectively. For convenience, we define the virtual AoDs and AoAs as
\begin{subequations}\label{eq_AoD_AoA}
	\begin{align}
		&\begin{aligned}&\vartheta_{k, j}^{\mathrm{t}} = \cos \theta_{k,j}^{\mathrm{t}} \cos \phi_{k,j}^{\mathrm{t}},~
		\varphi_{k, j}^{\mathrm{t}} = \cos \theta_{k,j}^{\mathrm{t}} \sin \phi_{k,j}^{\mathrm{t}},\\
		&\omega_{k, j}^{\mathrm{t}} = \sin \theta_{k,j}^{\mathrm{t}},~~~~~~~~~1 \leq k \leq K,~1 \leq j \leq L_{k}^{\mathrm{t}}, \end{aligned} \label{eq_AoD}\\
		&\begin{aligned}&\vartheta_{k, i}^{\mathrm{r}} = \cos \theta_{k,i}^{\mathrm{r}} \cos \phi_{k,i}^{\mathrm{r}},~ 
		\varphi_{k, i}^{\mathrm{r}} = \cos \theta_{k,i}^{\mathrm{r}} \sin \phi_{k,i}^{\mathrm{r}},~\\
		&\omega_{k, i}^{\mathrm{r}} = \sin \theta_{k,i}^{\mathrm{r}},~~~~~~~~~1 \leq k \leq K,~1 \leq i \leq L_{k}^{\mathrm{r}}. \end{aligned}  \label{eq_AoA}
	\end{align}
\end{subequations}

Denoting $\lambda$ as the carrier wavelength, the transmit and receive field-response vectors (FRVs) for the channel between user $k$ and the BS are obtained as \cite{zhu2022MAmodel,ma2022MAmimo}
\begin{subequations}\label{eq_field_res}
	\begin{align}
		&\mathbf{g}_{k}(\mathbf{u}_{k}) = \left[\mathrm{e}^{\mathrm{j}\frac{2\pi}{\lambda}\rho_{k,1}^{\mathrm{t}}(\mathbf{u}_{k})},
		\mathrm{e}^{\mathrm{j}\frac{2\pi}{\lambda}\rho_{k,2}^{\mathrm{t}}(\mathbf{u}_{k})},
		\cdots, \mathrm{e}^{\mathrm{j}\frac{2\pi}{\lambda}\rho_{k,L_{k}^{\mathrm{t}}}^{\mathrm{t}}(\mathbf{u}_{k})}\right]^{\mathrm{T}}, \label{eq_field_resTx}\\
		&\mathbf{f}_{k}(\mathbf{v}_{n}) = \left[\mathrm{e}^{\mathrm{j}\frac{2\pi}{\lambda}\rho_{k,1}^{\mathrm{r}}(\mathbf{v}_{n})},
		\mathrm{e}^{\mathrm{j}\frac{2\pi}{\lambda}\rho_{k,2}^{\mathrm{r}}(\mathbf{v}_{n})},
		\cdots, \mathrm{e}^{\mathrm{j}\frac{2\pi}{\lambda}\rho_{k,L_{k}^{\mathrm{r}}}^{\mathrm{r}}(\mathbf{v}_{n})}\right]^{\mathrm{T}}, \label{eq_field_resRx}
	\end{align}
\end{subequations}
with $1 \leq k \leq K$ and $1 \leq n \leq N$, where $\rho_{k,j}^{\mathrm{t}}(\mathbf{u}_{k})=x_{k} \vartheta_{k, j}^{\mathrm{t}}+y_{k} \varphi_{k, j}^{\mathrm{t}}+z_{k} \omega_{k, j}^{\mathrm{t}}$, $1 \leq j \leq L_{k}^{\mathrm{t}}$, represents the difference of the signal propagation distance for the $j$-th transmit channel path between MA position $\mathbf{u}_{k}$ and the origin (i.e., $O_{k}$) of the local coordinate system at user $k$. It indicates that the phase difference of the coefficient of the $j$-th transmit channel path for user $k$ between MA location $\mathbf{u}_{k}$ and $O_{k}$ is given by $\frac{2\pi}{\lambda}\rho_{k,j}^{\mathrm{t}}(\mathbf{u}_{k})$. Thus, the transmit FRV, $\mathbf{g}_{k}(\mathbf{u}_{k})$, characterizes the phase differences in all $L_{k}^{\mathrm{t}}$ transmit paths from user $k$ to the BS. Similarly, the receive FRV, $\mathbf{f}_{k}(\mathbf{v}_{n})$, accounts for the phase differences in all $L_{k}^{\mathrm{r}}$ receive paths from user $k$ to the BS, where $\rho_{k,i}^{\mathrm{r}}(\mathbf{v}_{n})=X_{n} \vartheta_{k, i}^{\mathrm{r}}+Y_{n} \varphi_{k, i}^{\mathrm{r}}+Z_{n} \omega_{k, i}^{\mathrm{r}}$, $1 \leq i \leq L_{k}^{\mathrm{r}}$, represents the difference of the signal propagation distance for the $i$-th receive channel path between BS-antenna position $\mathbf{v}_{n}$ and the origin of the local coordinate system at the BS.

Then, we define the path-response matrix (PRM), $\mathbf{\Sigma}_{k} \in \mathbb{C}^{L_{k}^{\mathrm{r}} \times L_{k}^{\mathrm{t}}}$, to represent the response between all the transmit and receive channel paths from $O_{k}$ to $O_{0}$, $1 \leq k \leq K$. Specifically, the entry in the $i$-th row and $j$-th column of $\mathbf{\Sigma}_{k}$ is the response coefficient between the $j$-th transmit path and the $i$-th receive path for user $k$. As a result, the channel vector between user $k$ and the BS can be expressed as \cite{zhu2022MAmodel,ma2022MAmimo}
\begin{equation}\label{eq_channel}
	\mathbf{h}_{k}(\mathbf{u}_{k}) = \mathbf{F}_{k}^{\mathrm{H}} \mathbf{\Sigma}_{k} \mathbf{g}_{k} (\mathbf{u}_{k}),~1 \leq k \leq K,
\end{equation}
where $\mathbf{F}_{k}=\left[\mathbf{f}_{k}(\mathbf{v}_{1}),\mathbf{f}_{k}(\mathbf{v}_{2}),\cdots,\mathbf{f}_{k}(\mathbf{v}_{N})\right] \in \mathbb{C}^{L_{k}^{\mathrm{r}} \times N}$ is the field-response matrix (FRM) at the BS, which is a constant matrix because the antennas of the BS have fixed positions. As can be observed from \eqref{eq_channel}, the positioning optimization of the MA for each user can change its FRV, which yields a varying linear combination of the columns in matrix $\mathbf{F}_{k}^{\mathrm{H}} \mathbf{\Sigma}_{k}$. Thus, the channel vector between each user and the BS can be significantly changed by moving the antenna of the user in a local region.

\subsection{Problem Formulation}
For the uplink transmission employing linear combining at the BS, the receive SINR of the signal from user $k$ is given by
\begin{equation}\label{eq_SINR}
	\gamma_{k} = \frac{\left|\mathbf{w}_{k}^{\mathrm{H}} \mathbf{h}_{k}(\mathbf{u}_{k})\right|^2 p_{k}}{\sum \limits_{q=1, q \neq k}^{K} \left|\mathbf{w}_{k}^{\mathrm{H}} \mathbf{h}_{q}(\mathbf{u}_{q})\right|^{2} p_{q}+\left\|\mathbf{w}_{k}\right\|_{2}^{2} \sigma^{2}},~1 \leq k \leq K,
\end{equation}
and thus the achievable rate for user $k$ is obtained as $R_{k} = \log_{2}\left(1+\gamma_{k}\right)$.

In this paper, we aim at minimizing the total transmit power of multiple users by jointly optimizing the position of MA for each user, the transmit power of each user, and the receive combining matrix of the BS, subject to a minimum-achievable-rate requirement for each user. Let $\mathbf{p}=[p_{1},p_{2},\cdots,p_{K}]^{\mathrm{T}}$ denote the vector of transmit power of the users. Accordingly, the optimization problem can be formulated as\footnote{The goal of this paper is to characterize the theoretical performance limit of the MA-enhanced MAC, where perfect FRI is assumed to be available at the BS for optimization, including the AoAs, AoDs, and PRMs of the channel paths for all users. The FRI acquisition for MA systems is beyond the scope of this paper and can be referred to \cite{ma2023MAestimation}. Nonetheless, the impact of imperfect FRI on the considered MA-enabled multiple access systems will be evaluated via simulations in this paper.}
\begin{subequations}\label{eq_problem}
	\begin{align}
		\mathop{\min}\limits_{\tilde{\mathbf{u}}, \mathbf{p}, \mathbf{W}}~~~
		&\sum \limits_{k=1}^{K} p_{k} \label{eq_problem_a}\\
		\mathrm{s.t.}~~~~ &\log_{2}\left(1+\gamma_{k}\right) \geq r_{k},~1 \leq k \leq K, \label{eq_problem_b}\\
		&\mathbf{u}_{k} \in \mathcal{C}_{k},~1 \leq k \leq K, \label{eq_problem_c}\\
		&p_{k} \geq 0,~1 \leq k \leq K, \label{eq_problem_d}
	\end{align}
\end{subequations}
where constraint \eqref{eq_problem_b} indicates that the achievable rate for user $k$ should be no smaller than its minimum requirement $r_{k}$. Constraint \eqref{eq_problem_c} confines that the MA of user $k$ is located in its moving region, $\mathcal{C}_{k}$. Constraint \eqref{eq_problem_d} guarantees that the transmit power of each user is non-negative. Problem \eqref{eq_problem} is difficult to solve because the channel vectors and achievable rates of the users are highly non-convex w.r.t. the positions of MAs. Besides, the coupling between these high-dimensional matrix/vector variables makes this problem more intractable. Problem \eqref{eq_problem} cannot be optimally solved with existing optimization tools in polynomial time. Thus, we develop suboptimal solutions for \eqref{eq_problem} in the next section.

\section{Proposed Solution}
Due to the coupling between the positions of MAs, the transmit power of users, and the receive combining matrix of the BS, problem \eqref{eq_problem} cannot be optimally solved efficiently. In general, the alternating optimization among the three matrix/vector variables may lead to a (undesired) local optimum. For example, given the optimal receive combining matrix of the BS designed based on the channel vectors between the BS and users, the optimized positions of MAs cannot be significantly changed compared to those in the last iteration. This is because the channel vectors of other MA positions do not match the receive combining matrix and are highly likely to decrease the effective channel gains of the target signals as well as increase the interference among multiple users. To address this issue, we propose to leverage the ZF and MMSE combining methods for expressing the receive combining matrix as a function of the MA positioning vector, and then optimize the positions of MAs for minimizing the total transmit power of multiple users. The two solutions based on ZF and MMSE combining are presented in the following, respectively.

\subsection{ZF-Based Solution}
For any given positions of MAs, $\tilde{\mathbf{u}}$, the channel vectors between the BS and users are fixed\footnote{In this paper, we assume that the MAC matrix for multiple users is of column full rank. Otherwise, the minimum-achievable-rate requirements for the users are difficult to fulfill due to channel correlation. For the case where the channel matrix is not of column full rank, we can decrease the number of served users and formulate a similar problem.}. Thus, the ZF combining matrix can be expressed as a function of the MA positioning vector \cite{TseFundaWC}, i.e.,
\begin{equation}\label{eq_ZF}
	\begin{aligned}
	\mathbf{W}_{\mathrm{ZF}}(\tilde{\mathbf{u}})&=\left(\mathbf{H}(\tilde{\mathbf{u}})^{\mathrm{H}}\right)^{\dagger}=\mathbf{H}(\tilde{\mathbf{u}})\left(\mathbf{H}(\tilde{\mathbf{u}})^{\mathrm{H}}\mathbf{H}(\tilde{\mathbf{u}})\right)^{-1}\\
	&\triangleq \left[\bar{\mathbf{w}}_{1},\bar{\mathbf{w}}_{2},\cdots,\bar{\mathbf{w}}_{K}\right].
	\end{aligned}
\end{equation}
Substituting \eqref{eq_ZF} into \eqref{eq_SINR}, the receive SINR of the signal from user $k$, $1 \leq k \leq K$, can be rewritten as
\begin{equation}\label{eq_SINR_ZF}
	\bar{\gamma}_{k}=\frac{p_{k}}{\left\|\bar{\mathbf{w}}_{k}\right\|_{2}^{2} \sigma^{2}}= \frac{p_{k}}{\left\|\mathbf{H}(\tilde{\mathbf{u}})\left[\left(\mathbf{H}(\tilde{\mathbf{u}})^{\mathrm{H}} \mathbf{H}(\tilde{\mathbf{u}})\right)^{-1}\right]_{:, k}\right\|_{2}^{2} \sigma^2}.
\end{equation}
Furthermore, we substitute \eqref{eq_SINR_ZF} into \eqref{eq_problem_b} and obtain the minimum transmit power of user $k$ for satisfying the achievable-rate requirement as
\begin{equation}\label{eq_power_ZF}
	\begin{aligned}
		&\log_{2}\left(1+\bar{\gamma}_{k}\right) \geq r_{k}\\
		\Rightarrow &\frac{p_{k}}{\left\|\mathbf{H}(\tilde{\mathbf{u}})\left[\left(\mathbf{H}(\tilde{\mathbf{u}})^{\mathrm{H}} \mathbf{H}(\tilde{\mathbf{u}})\right)^{-1}\right]_{:, k}\right\|_{2}^{2} \sigma^2} \geq 2^{r_{k}}-1\\
		\Rightarrow &p_{k} \geq \left\|\mathbf{H}(\tilde{\mathbf{u}})\left[\left(\mathbf{H}(\tilde{\mathbf{u}})^{\mathrm{H}} \mathbf{H}(\tilde{\mathbf{u}})\right)^{-1}\right]_{:, k}\right\|_{2}^{2} \eta_{k} \sigma^{2} \triangleq \bar{p}_{k},
	\end{aligned}
\end{equation}
where $\eta_{k}=2^{r_{k}}-1$ represents the minimum-SINR requirement for user $k$, $1 \leq k \leq K$. Besides, to minimize the total transmit power under any given MA positioning vector $\tilde{\mathbf{u}}$, the optimal transmit power of each user should be set as its lower bound, i.e., $\bar{p}_{k}$ in \eqref{eq_power_ZF}, for satisfying the minimum-achievable-rate requirement. Note that the transmit power $\bar{p}_{k}$ is always non-negative, and thus constraint \eqref{eq_problem_d} is satisfied. As a result, under any given $\tilde{\mathbf{u}}$ and the corresponding ZF combining matrix, the optimal solution for the total transmit power of users for satisfying the minimum-achievable-rate constraint can be expressed as
\begin{equation}\label{eq_total_power_ZF}
	\begin{aligned}
	\sum \limits _{k=1}^{K} \bar{p}_{k} &= \sum \limits _{k=1}^{K} \left\|\mathbf{H}(\tilde{\mathbf{u}})\left[\left(\mathbf{H}(\tilde{\mathbf{u}})^{\mathrm{H}} \mathbf{H}(\tilde{\mathbf{u}})\right)^{-1}\right]_{:, k}\right\|_{2}^{2} \eta_{k} \sigma^{2} \\
	&= \left\|\mathbf{H}(\tilde{\mathbf{u}})\left(\mathbf{H}(\tilde{\mathbf{u}})^{\mathrm{H}} \mathbf{H}(\tilde{\mathbf{u}})\right)^{-1}\mathbf{\Omega}^{1/2}\right\|_{\mathrm{F}}^{2} \\
	&= \operatorname{tr}\left\{ \left(\mathbf{H}(\tilde{\mathbf{u}})^{\mathrm{H}} \mathbf{H}(\tilde{\mathbf{u}})\right)^{-1}\mathbf{\Omega} \right\}\\
	&= \operatorname{tr}\left\{ \left(\mathbf{\Omega}^{-1}\mathbf{H}(\tilde{\mathbf{u}})^{\mathrm{H}} \mathbf{H}(\tilde{\mathbf{u}})\right)^{-1} \right\}\\
	&= \sum_{k=1}^K \frac{1}{\lambda_k\left\{\mathbf{\Omega}^{-1} \mathbf{H}(\tilde{\mathbf{u}})^{\mathrm{H}} \mathbf{H}(\tilde{\mathbf{u}})\right\}} \triangleq g(\tilde{\mathbf{u}}),
	\end{aligned}
\end{equation}
where $\mathbf{\Omega}=\operatorname{diag}\left\{\eta_{1} \sigma^{2}, \eta_{2} \sigma^{2}, \cdots, \eta_{K} \sigma^{2}\right\}$ is a diagonal matrix with constant elements.\\ $\lambda_k\left\{\mathbf{\Omega}^{-1} \mathbf{H}(\tilde{\mathbf{u}})^{\mathrm{H}} \mathbf{H}(\tilde{\mathbf{u}})\right\}$ denotes the $k$-th eigenvalue of matrix $\mathbf{\Omega}^{-1} \mathbf{H}(\tilde{\mathbf{u}})^{\mathrm{H}} \mathbf{H}(\tilde{\mathbf{u}})$. Then, problem \eqref{eq_problem} can be transformed into\footnote{Since ZF combining at the BS is not optimal, problem \eqref{eq_problem_ZF} is not equivalent to problem \eqref{eq_problem} but provides a suboptimal solution for it.}
\begin{subequations}\label{eq_problem_ZF}
	\begin{align}
		\mathop{\min}\limits_{\tilde{\mathbf{u}}}~~~
		&g(\tilde{\mathbf{u}}) \label{eq_problem_ZF_a}\\
		\mathrm{s.t.}~~~~ &\mathbf{u}_{k} \in \mathcal{C}_{k},~1 \leq k \leq K. \label{eq_problem_ZF_b}
	\end{align}
\end{subequations}

Since the objective function, i.e., $g(\tilde{\mathbf{u}})$ defined in \eqref{eq_total_power_ZF}, is highly non-convex w.r.t. $\tilde{\mathbf{u}}$, it is challenging to obtain the globally optimal solution for problem \eqref{eq_problem_ZF}. Next, we propose the MDD framework for optimizing the MA positioning vector. Specifically, the initial MA positioning vector is set as $\tilde{\mathbf{u}}^{(0)}=\mathbf{0}_{3K}$, which means that the MA of each user is initially placed at the origin of its local coordinate system. Let $\tilde{\mathbf{u}}^{(t-1)}$ denote the MA positioning vector obtained in the $(t-1)$-th iteration. Then, the MA positioning vector in the $t$-th iteration is updated by moving along $\bar{M}$ candidate descent directions, i.e.,
\begin{equation}\label{eq_position_ZF_candi}
	\tilde{\mathbf{u}}^{(t)}_{m}=\mathcal{B}\left\{\tilde{\mathbf{u}}^{(t-1)}+\bar{\tau}^{(t)}_{m} \bar{\mathbf{d}}^{(t)}_{m}\right\}, 1 \leq m \leq \bar{M},
\end{equation}
where $\bar{\mathbf{d}}^{(t)}_{m}$ and $\bar{\tau}^{(t)}_{m}$ are the $m$-th candidate descent direction and the corresponding step size in the $t$-th iteration, which will be specified later. $\mathcal{B}\left\{\tilde{\mathbf{u}}\right\}$ is a function which projects each element in $\tilde{\mathbf{u}}$ to the nearest boundary of its feasible region if this element exceeds the feasible region, i.e.,
\begin{equation}\label{eq_project}
	\left[\mathcal{B}\left\{\tilde{\mathbf{u}}\right\}\right]_{i}=\left\{
	\begin{aligned}
		&\left[\tilde{\mathbf{u}}\right]_{i}^{\min}, ~\text{if } \left[\tilde{\mathbf{u}}\right]_{i} < \left[\tilde{\mathbf{u}}\right]_{i}^{\min},\\
		&\left[\tilde{\mathbf{u}}\right]_{i}, ~~~~\text{if } \left[\tilde{\mathbf{u}}\right]_{i}^{\min} \leq \left[\tilde{\mathbf{u}}\right]_{i} \leq \left[\tilde{\mathbf{u}}\right]_{i}^{\max},\\
		&\left[\tilde{\mathbf{u}}\right]_{i}^{\max}, ~\text{if } \left[\tilde{\mathbf{u}}\right]_{i} > \left[\tilde{\mathbf{u}}\right]_{i}^{\max},
	\end{aligned}\right.
\end{equation}
where $\left[\tilde{\mathbf{u}}\right]_{i}^{\min}$ and $\left[\tilde{\mathbf{u}}\right]_{i}^{\max}$ denote the lower-bound and upper-bound on the feasible region of the $i$-th element of $\tilde{\mathbf{u}}$, respectively. According to the definition $\tilde{\mathbf{u}}=\left[\mathbf{u}_{1}^{\mathrm{T}},\mathbf{u}_{2}^{\mathrm{T}},\cdots,\mathbf{u}_{K}^{\mathrm{T}}\right]^{\mathrm{T}}$, the $i$-th element of $\tilde{\mathbf{u}}$ corresponds to the $q_{i}$-th element of $\mathbf{u}_{k_{i}}$, with $k_{i}=\lfloor i/3 \rfloor+1$ and $q_{i}=i-3(k_{i}-1)$. Thus, we have $\left[\tilde{\mathbf{u}}\right]_{i}^{\min}=x_{k_{i}}^{\min}$ and $\left[\tilde{\mathbf{u}}\right]_{i}^{\max}=x_{k_{i}}^{\max}$ for $q_{i}=1$, $\left[\tilde{\mathbf{u}}\right]_{i}^{\min}=y_{k_{i}}^{\min}$ and $\left[\tilde{\mathbf{u}}\right]_{i}^{\max}=y_{k_{i}}^{\max}$ for $q_{i}=2$, and $\left[\tilde{\mathbf{u}}\right]_{i}^{\min}=z_{k_{i}}^{\min}$ and $\left[\tilde{\mathbf{u}}\right]_{i}^{\max}=z_{k_{i}}^{\max}$ for $q_{i}=3$. The utilization of projection function  $\mathcal{B}\left\{\tilde{\mathbf{u}}\right\}$ is to ensure that the solution for MA positioning is always located in the feasible region during the iterations.

The $\bar{M}$ candidate descent directions in \eqref{eq_position_ZF_candi} can be generated by the weighted combinations of $\bar{Q}$ descent-direction components, i.e., 
\begin{equation}\label{eq_position_ZF_MDD}
	\bar{\mathbf{d}}^{(t)}_{m} = \bar{\mathbf{\Psi}}^{(t)} \bar{\mathbf{c}}_{m},~1 \leq m \leq \bar{M},
\end{equation}
where each column of $\bar{\mathbf{\Psi}}^{(t)} \in \mathbb{R}^{3K \times \bar{Q}}$ represents a descent-direction component, which can be obtained by adopting classical methods, such as the opposite direction of gradient, the opposite of gradient with momentum, the descent direction of the (quasi-) Newton's method, etc \cite{chong2013introduction}. Constant vector $\bar{\mathbf{c}}_{m} \in  \mathbb{R}^{\bar{Q} \times 1}$, $1 \leq m \leq \bar{M}$, represents the combination weights for the $\bar{Q}$ component descent directions for generating the $m$-th candidate descent direction. In general, the descent-direction components in $\bar{\mathbf{\Psi}}^{(t)}$ may not always be the optimal direction for decreasing the objective function but imply the significant information of directions that may potentially decrease it. In other words, both the $\bar{Q}$ descent-direction components themselves and their weighted combinations can serve as candidate directions for decreasing the objective function. Thus, the combination weight vectors $\{\bar{\mathbf{c}}_{m}\}$ are introduced to control the generation of candidate descent directions. It is worth noting that the conventional gradient descent and Newton's methods can both be regarded as a special case of the proposed MDD framework by only setting one candidate descent direction. Similar methodologies have also been adopted in the existing literature on optimization, e.g., \cite{streeter2023universal}.

In addition, the step size for each candidate descent direction significantly impacts the performance of the MDD algorithm. In this paper, we employ the backtracking line search to obtain an appropriate step size \cite{boyd2004convex}. Specifically, for the $m$-th candidate descent direction in the $t$-th iteration, we start with a large positive step size, $\bar{\tau}^{(t)}_{m}=\bar{\tau}$, and repeatedly shrink it by a factor $\bar{\kappa} \in (0,1)$, i.e., $\bar{\tau}^{(t)}_{m} \leftarrow \bar{\kappa}\bar{\tau}^{(t)}_{m}$, until the Armijo–Goldstein condition is satisfied, i.e.,
\begin{equation}\label{eq_Arm_ZF}
	\begin{aligned}
	&g(\tilde{\mathbf{u}}^{(t)}_{m}) \leq g(\tilde{\mathbf{u}}^{(t-1)})\\
	&~~~~~~~~~~-\bar{\xi} \bar{\tau}^{(t)}_{m} \left| \nabla_{\tilde{\mathbf{u}}} g\left(\tilde{\mathbf{u}}^{(t-1)}\right)^{\mathrm{T}} \bar{\mathbf{d}}^{(t)}_{m} \right|,~1 \leq m \leq \bar{M},		
\end{aligned}
\end{equation}
where $\nabla_{\tilde{\mathbf{u}}} g\left(\tilde{\mathbf{u}}^{(t-1)}\right)$ denotes the gradient of function $g\left(\tilde{\mathbf{u}}\right)$ at location $\tilde{\mathbf{u}}^{(t-1)}$ and $\bar{\xi} \in (0,1)$ is a given control parameter to evaluate whether the current step size achieves an adequate decrease in the objective function. 

Finally, the MA positioning vector in the $t$-th iteration is selected from $\bar{M}$ candidate solutions in \eqref{eq_position_ZF_candi} which can achieve the minimum objective value in \eqref{eq_problem_ZF_a}, i.e.,
\begin{equation}\label{eq_position_ZF}
	\tilde{\mathbf{u}}^{(t)}=\arg \min \limits_{{\{\tilde{\mathbf{u}}^{(t)}_{m}\}}_{1 \leq m \leq \bar{M}}} g(\tilde{\mathbf{u}}^{(t)}_{m}).
\end{equation}
The overall algorithm terminates until the decrement on the objective value in \eqref{eq_problem_a} is below a small positive value, $\bar{\epsilon}$.

\begin{algorithm}[t]\small
	\caption{The ZF-based MDD framework for solving problem \eqref{eq_problem}.}
	\label{alg_ZF}
	\begin{algorithmic}[1]
		\REQUIRE ~$N$, $K$, $\sigma^{2}$, $\lambda$, $\{\mathcal{C}_{k}\}$, $\{r_{k}\}$, $\{\mathbf{\Sigma}_{k}\}$, $\{\mathbf{F}_{k}\}$, $\{\theta_{k,j}^{\mathrm{t}}\}$, $\{\phi_{k,j}^{\mathrm{t}}\}$, $\bar{M}$, $\bar{Q}$, $\{\bar{\mathbf{c}}_{m}\}$, $\bar{\tau}$, $\bar{\xi}$, $\bar{\kappa}$, $\bar{\epsilon}$, $\bar{T}_{\max}$.
		\ENSURE ~$\tilde{\mathbf{u}}_{\mathrm{ZF}}$, $\mathbf{W}_{\mathrm{ZF}}$, $\mathbf{p}_{\mathrm{ZF}}$. \\
		\STATE Initialize $\tilde{\mathbf{u}}^{(0)}=\mathbf{0}_{3K}$.
		\STATE Obtain $g(\tilde{\mathbf{u}}^{(0)})$ according to \eqref{eq_total_power_ZF}.
		\FOR   {$t=1:1:\bar{T}_{\max}$}
		\STATE Calculate descent-direction components $\bar{\mathbf{\Psi}}^{(t)}$.
		\FOR   {$m=1:1:\bar{M}$}
		\STATE Calculate candidate descent direction $\bar{\mathbf{d}}^{(t)}_{m}$ in \eqref{eq_position_ZF_MDD}.
		\STATE Initialize step size $\bar{\tau}^{(t)}_{m}=\bar{\tau}$.
		\STATE Update candidate position vector $\tilde{\mathbf{u}}^{(t)}_{m}$ in \eqref{eq_position_ZF_candi}.
		\WHILE {\eqref{eq_Arm_ZF} is not satisfied}
		\STATE Shrink the step size $\bar{\tau}^{(t)}_{m}\leftarrow \bar{\kappa}\bar{\tau}^{(t)}_{m}$.
		\STATE Update candidate position vector $\tilde{\mathbf{u}}^{(t)}_{m}$ in \eqref{eq_position_ZF_candi}.
		\ENDWHILE
		\ENDFOR 
		\STATE Update MA positioning vector $\tilde{\mathbf{u}}^{(t)}$ according to \eqref{eq_position_ZF}.
		\IF    {$\left|g(\tilde{\mathbf{u}}^{(t)})-g(\tilde{\mathbf{u}}^{(t-1)})\right|<\bar{\epsilon}$}
		\STATE Break.
		\ENDIF
		\ENDFOR
		\STATE Set the MA positioning vector as $\tilde{\mathbf{u}}_{\mathrm{ZF}}=\tilde{\mathbf{u}}^{(t)}$.
		\STATE Calculate the ZF combining matrix $\mathbf{W}_{\mathrm{ZF}}$ according to \eqref{eq_ZF}.
		\STATE Calculate the transmit power $\mathbf{p}_{\mathrm{ZF}}$ according to \eqref{eq_power_ZF}.
		\RETURN $\tilde{\mathbf{u}}_{\mathrm{ZF}}$, $\mathbf{W}_{\mathrm{ZF}}$, $\mathbf{p}_{\mathrm{ZF}}$.
	\end{algorithmic}
\end{algorithm}

The ZF-based MDD framework for solving problem \eqref{eq_problem} is summarized in Algorithm \ref{alg_ZF}, where $\bar{T}_{\max}$ denotes the maximum number of iterations. In lines 3-18, the MA positioning vector is iteratively updated by using the MDD method, where the step size for each candidate descent direction in each iteration is obtained by backtracking line search in lines 9-12. The convergence of Algorithm \ref{alg_ZF} is analyzed as follows. On one hand, due to the fact of $\bar{\tau}^{(t)}_{m}>0$, $\bar{\xi}>0$, and $\left| \nabla_{\tilde{\mathbf{u}}} g\left(\tilde{\mathbf{u}}^{(t-1)}\right)^{\mathrm{T}} \bar{\mathbf{d}}^{(t)}_{m} \right| \geq 0$, we always have $g(\tilde{\mathbf{u}}^{(t)}) \leq g(\tilde{\mathbf{u}}^{(t)}_{m}) \leq g(\tilde{\mathbf{u}}^{(t-1)})$ during the iterations. On the other hand, for the case of $\left| \nabla_{\tilde{\mathbf{u}}} g\left(\tilde{\mathbf{u}}^{(t-1)}\right)^{\mathrm{T}} \bar{\mathbf{d}}^{(t)}_{m} \right| > 0$, we can always find a sufficiently small positive $\bar{\tau}^{(t)}_{m}$ satisfying $g(\tilde{\mathbf{u}}^{(t)}) \leq g(\tilde{\mathbf{u}}^{(t)}_{m}) < g(\tilde{\mathbf{u}}^{(t-1)})$, unless all elements of $\tilde{\mathbf{u}}^{(t-1)}$ with nonzero partial derivatives are located on the boundary of the feasible region as well as the corresponding descent direction is towards the outside of the feasible region. Thus, Algorithm \ref{alg_ZF} terminates at a solution with all candidate descent directions equal to zero or a solution located on the boundary of the feasible region defined by \eqref{eq_problem_ZF_b}, which yields a local optimum for problem \eqref{eq_problem}. To improve the capability of global search, the initial step size, $\bar{\tau}$, should be set sufficiently large such that the solution can get rid of local optimum. 

In Algorithm \ref{alg_ZF}, the maximum computational complexity for calculating the descent-direction components based on first-order methods in line 4 is $\mathcal{O}(K^{3}N\bar{Q})$. Denoting the maximum number of iterations for backtracking line search in lines 9-12 as $\bar{I}_{\max}$, the total computational complexity for searching all candidate descent directions is given by $\mathcal{O}(\bar{I}_{\max}K^{2}N\bar{M})$. As a result, the maximum computational complexity of Algorithm \ref{alg_ZF} for solving problem \eqref{eq_problem} is $\mathcal{O}\left(\bar{T}_{\max}(K^{3}N\bar{Q}+\bar{I}_{\max}K^{2}N\bar{M})\right)$.

\subsection{MMSE-Based Solution}
For any given MA positioning vector (i.e., $\tilde{\mathbf{u}}$) and transmit power of the users (i.e., $\mathbf{p}$), the optimal linear combining matrix of the BS for maximizing the achievable rate region is given by the MMSE receiver \cite{TseFundaWC}. In other words, subject to the minimum-achievable-rate requirement of each user, the MMSE combining matrix is optimal for minimizing the total transmit power and given by
\begin{equation}\label{eq_MMSE}
	\begin{aligned}
		\mathbf{W}_{\mathrm{MMSE}}(\tilde{\mathbf{u}},\mathbf{P})&=\left(\mathbf{H}(\tilde{\mathbf{u}}) \mathbf{P} \mathbf{H}(\tilde{\mathbf{u}})^{\mathrm{H}} + \sigma^{2} \mathbf{I}_{N}\right)^{-1}\mathbf{H}(\tilde{\mathbf{u}})\\
		&\triangleq \left[\hat{\mathbf{w}}_{1},\hat{\mathbf{w}}_{2},\cdots,\hat{\mathbf{w}}_{K}\right],
	\end{aligned}
\end{equation}
with $\hat{\mathbf{w}}_{k}=\left(\mathbf{H}(\tilde{\mathbf{u}}) \mathbf{P} \mathbf{H}(\tilde{\mathbf{u}})^{\mathrm{H}} + \sigma^{2} \mathbf{I}_{N}\right)^{-1} \mathbf{h}_{k}(\mathbf{u}_{k})$, $1 \leq k \leq K$, and $\mathbf{P}=\operatorname{diag}\{\mathbf{p}\}$. Substituting \eqref{eq_MMSE} into \eqref{eq_SINR}, the receive SINR of the signal from user $k$ can be rewritten as
\begin{equation}\label{eq_SINR_MMSE}
	\begin{aligned}
		\hat{\gamma}_{k}&=\frac{\left|\hat{\mathbf{w}}_{k}^{\mathrm{H}} \mathbf{h}_{k}(\mathbf{u}_{k})\right|^2 p_{k}}{\sum \limits_{q=1, q \neq k}^{K} \left|\hat{\mathbf{w}}_{k}^{\mathrm{H}} \mathbf{h}_{q}(\mathbf{u}_q)\right|^{2} p_{q}+\left\|\hat{\mathbf{w}}_{k}\right\|_{2}^{2} \sigma^{2}}\\
		&\triangleq \frac{A_{k,k} p_{k}}{\sum \limits_{q=1, q \neq k}^{K} A_{k,q} p_{q}+b_{k}},~1 \leq k \leq K,
	\end{aligned}
\end{equation}
with intermediate variables $A_{k,q} \triangleq \left|\hat{\mathbf{w}}_{k}^{\mathrm{H}} \mathbf{h}_{q}(\mathbf{u}_q)\right|^{2}=\left|\mathbf{h}_{k}(\mathbf{u}_{k})^{\mathrm{H}} \left(\mathbf{H}(\tilde{\mathbf{u}}) \mathbf{P} \mathbf{H}(\tilde{\mathbf{u}})^{\mathrm{H}} + \sigma^{2} \mathbf{I}_{N}\right)^{-1}  \mathbf{h}_{q}(\mathbf{u}_{q})\right|^{2}$ and $b_{k} \triangleq \left\|\hat{\mathbf{w}}_{k}\right\|_{2}^{2} \sigma^{2} = \left\|\left(\mathbf{H}(\tilde{\mathbf{u}}) \mathbf{P} \mathbf{H}(\tilde{\mathbf{u}})^{\mathrm{H}} + \sigma^{2} \mathbf{I}_{N}\right)^{-1} \mathbf{h}_{k}(\mathbf{u}_{k})\right\|_{2}^{2} \sigma^{2}$, $1 \leq k \leq K$ and $1 \leq q \leq K$. 

Let $\mathbf{A} \in \mathbb{R}^{K \times K}$ denote a matrix with the entry in the $k$-th row and $q$-th column given by $A_{k,q}$, and $\mathbf{b} \in \mathbb{R}^{K \times 1}$ denote a vector with the $k$-th entry given by $b_{k}$. It is easy to verify that to minimize the total transmit power, the SINR of each user should be exactly equal to its minimum requirement, i.e., $\hat{\gamma}_{k} = \eta_{k}$ \cite{Boche2002duality}, which can be equivalently expressed as $A_{k,k}/\eta_{k} \times p_{k} = \sum_{q=1, q \neq k}^{K} A_{k,q} p_{q}+b_{k}$,  $1 \leq k \leq K$. The matrix form of these equations w.r.t. $\mathbf{p}$ is given by
\begin{equation}\label{eq_powerEq_MMSE}
	\begin{aligned}
		\left(\mathbf{D}-\mathbf{\Psi}\right)\mathbf{p}=\mathbf{b},
	\end{aligned}
\end{equation}
with $\mathbf{D}=\operatorname{diag}\{A_{1,1}/\eta_{1},A_{2,2}/\eta_{2},\cdots,A_{K,K}/\eta_{K}\}$, $\left[\mathbf{\Psi}\right]_{k,q}=A_{k,q}$ for $1 \leq k \neq q \leq K$ and $\left[\mathbf{\Psi}\right]_{k,k}=0$ for $1 \leq k \leq K$. Thus, the optimal solution for the transmit power is obtained as
\begin{equation}\label{eq_power_MMSE}
	\begin{aligned}
		\hat{\mathbf{p}}=\left(\mathbf{D}-\mathbf{\Psi}\right)^{-1}\mathbf{b}.
	\end{aligned}
\end{equation}
The total transmit power of users for satisfying the minimum-achievable-rate constraint can be expressed as
\begin{equation}\label{eq_total_power_MMSE}
	\begin{aligned}
		\sum \limits _{k=1}^{K} \hat{p}_{k} = \left\|\left(\mathbf{D}-\mathbf{\Psi}\right)^{-1}\mathbf{b}\right\|_{1} \triangleq f(\tilde{\mathbf{u}},\mathbf{P}).
	\end{aligned}
\end{equation}
Note that an implicit constraint should be considered to ensure that the transmit power of each user is non-negative, i.e., $\hat{p}_{k} \geq 0$, $1 \leq k \leq K$. It has been shown in \cite{Boche2002duality} that the non-negative power constraint is equivalent to that the spectral radius of matrix $\mathbf{D}\mathbf{\Psi}^{-1}$ is smaller than 1. Then, problem \eqref{eq_problem} can be transformed into\footnote{Since MMSE combining at the BS is optimal, problem \eqref{eq_problem_MMSE} is equivalent to problem \eqref{eq_problem}.}
\begin{subequations}\label{eq_problem_MMSE}
	\begin{align}
		\mathop{\min}\limits_{\tilde{\mathbf{u}} ,\mathbf{P}}~~~
		&f(\tilde{\mathbf{u}},\mathbf{P}) \label{eq_problem_MMSE_a}\\
		\mathrm{s.t.}~~~~ &\mathbf{u}_{k} \in \mathcal{C}_{k},~1 \leq k \leq K, \label{eq_problem_MMSE_b}\\
		&\rho\left\{\mathbf{D}\mathbf{\Psi}^{-1}\right\}<1, \label{eq_problem_MMSE_c}
	\end{align}
\end{subequations}
where $\rho\left\{\mathbf{D}\mathbf{\Psi}^{-1}\right\}$ represents the spectral radius of matrix $\mathbf{D}\mathbf{\Psi}^{-1}$, which is equal to the maximum of the absolute values of its eigenvalues. 

Problem \eqref{eq_problem_MMSE} is more challenging to solve compared to \eqref{eq_problem_ZF} due to the following reasons. On one hand, constraint \eqref{eq_problem_MMSE_c} is non-convex w.r.t. the optimization variables. On the other hand, the MA positioning vector and the transmit power matrix are highly coupled in the objective function of \eqref{eq_problem_MMSE_a}. To tackle this non-convex problem, we propose an iterative algorithm to alternately update $\tilde{\mathbf{u}}$ and $\mathbf{P}$. Specifically, the MA positioning vector is initialized as $\tilde{\mathbf{u}}^{(0)}=\mathbf{0}_{3K}$. Under the given MA positioning, the initial transmit power is $\mathbf{P}^{(0)}$ obtained by alternately optimizing the MMSE combining matrix of the BS according to \eqref{eq_MMSE} and the transmit power of users according to \eqref{eq_power_MMSE} until the total transmit power is minimized. Let $\tilde{\mathbf{u}}^{(t-1)}$ and $\mathbf{P}^{(t-1)}$ denote the solution for MA positioning and transmit power in the $(t-1)$-th iteration, respectively. Then, the MA positioning vector in the $t$-th iteration is updated by using the proposed MDD method similar to that in Algorithm \ref{alg_ZF}, where $\hat{M}$ candidate descent directions are employed for updating the MA positioning vector, i.e.,
\begin{equation}\label{eq_position_MMSE_candi}
	\tilde{\mathbf{u}}^{(t)}_{m}=\mathcal{B}\left\{\tilde{\mathbf{u}}^{(t-1)}+\hat{\tau}^{(t)}_{m} \hat{\mathbf{d}}^{(t)}_{m}\right\}, 1 \leq m \leq \hat{M},
\end{equation}
where $\hat{\mathbf{d}}^{(t)}_{m}$ and $\hat{\tau}^{(t)}_{m}$ are the $m$-th candidate descent direction and the corresponding step size in the $t$-th iteration.

Similarly to Section III-A, the $\hat{M}$ candidate descent directions are generated by 
\begin{equation}\label{eq_position_MMSE_MDD}
	\hat{\mathbf{d}}^{(t)}_{m} = \hat{\mathbf{\Psi}}^{(t)} \hat{\mathbf{c}}_{m},~1 \leq m \leq \hat{M},
\end{equation}
where $\hat{\mathbf{\Psi}}^{(t)} \in \mathbb{R}^{3K \times \hat{Q}}$ comprises $\hat{Q}$ descent-direction components and $\hat{\mathbf{c}}_{m} \in  \mathbb{R}^{\hat{Q} \times 1}$, $1 \leq m \leq \hat{M}$ is the combination weights for generating the $m$-th candidate descent direction.

The step size for each candidate descent direction is also selected by backtracking line search such that the Armijo–Goldstein condition and the spectral radius constraint are both satisfied, i.e, 
\begin{subequations}\label{eq_Arm_MMSE}
	\begin{align}
		&\begin{aligned}&f\left(\tilde{\mathbf{u}}^{(t)}_{m},\mathbf{P}^{(t-1)}\right) \leq f\left(\tilde{\mathbf{u}}^{(t-1)},\mathbf{P}^{(t-1)}\right)\\
			&~~~~~~~~~~~~~~~~~ - \hat{\xi} \hat{\tau}^{(t)}_{m} \left|\nabla_{\tilde{\mathbf{u}}} f\left(\tilde{\mathbf{u}}^{(t-1)},\mathbf{P}^{(t-1)}\right)^{\mathrm{T}} \hat{\mathbf{d}}^{(t)}_{m} \right|,\end{aligned}\\
		&\rho\left\{\mathbf{D}^{(t)}_{m}(\mathbf{\Psi}^{(t)}_{m})^{-1}\right\}<1,~1 \leq m \leq \hat{M}.
	\end{align}
\end{subequations}

Subsequently, the MA positioning vector in the $t$-th iteration is selected from $\hat{M}$ candidate solutions in \eqref{eq_position_MMSE_candi} which can achieve the minimum objective value in \eqref{eq_problem_MMSE_a}, i.e.,
\begin{equation}\label{eq_position_MMSE}
	\tilde{\mathbf{u}}^{(t)}=\arg \min \limits_{{\{\tilde{\mathbf{u}}^{(t)}_{m}\}}_{1 \leq m \leq \hat{M}}} f\left(\tilde{\mathbf{u}}^{(t)}_{m},\mathbf{P}^{(t-1)}\right).
\end{equation}
Finally, we calculate the values of $\mathbf{D}^{(t)}$, $\mathbf{\Psi}^{(t)}$, $\mathbf{b}^{(t)}$, and $\hat{\mathbf{p}}^{(t)}$ according to \eqref{eq_power_MMSE}. The transmit power matrix in the $t$-th iteration is thus updated as
\begin{equation}\label{eq_power_MMSE_it}
	\begin{aligned}
		\mathbf{P}^{(t)} = \operatorname{diag}\{\hat{\mathbf{p}}^{(t)}\}.
	\end{aligned}
\end{equation}
The overall algorithm terminates until the decrement on the objective value in \eqref{eq_problem_a} is below a small positive value, $\hat{\epsilon}$.

\begin{algorithm}[t]\small
	\caption{The MMSE-based MDD framework for solving problem \eqref{eq_problem}.}
	\label{alg_MMSE}
	\begin{algorithmic}[1]
		\REQUIRE ~$N$, $K$, $\sigma^{2}$, $\lambda$, $\{\mathcal{C}_{k}\}$ $\{r_{k}\}$, $\{\mathbf{\Sigma}_{k}\}$, $\{\mathbf{F}_{k}\}$, $\{\theta_{k,j}^{\mathrm{t}}\}$, $\{\phi_{k,j}^{\mathrm{t}}\}$, $\hat{M}$, $\hat{Q}$, $\{\hat{\mathbf{c}}_{m}\}$, $\hat{\tau}$, $\hat{\xi}$, $\hat{\kappa}$, $\hat{\epsilon}$, $\hat{T}_{\max}$.
		\ENSURE ~$\tilde{\mathbf{u}}_{\mathrm{MMSE}}$, $\mathbf{W}_{\mathrm{MMSE}}$, $\mathbf{p}_{\mathrm{MMSE}}$. \\
		\STATE Initialize $\tilde{\mathbf{u}}^{(0)}=\mathbf{0}_{3K}$ and $\mathbf{p}$ according to \eqref{eq_power_ZF}.
		\WHILE {The decrement on $\sum _{k=1}^{K} p_{k}$ is larger than $\hat{\epsilon}$}
		\STATE Update $\mathbf{P}=\operatorname{diag}\{\mathbf{p}\}$.
		\STATE Update $\mathbf{W}_{\mathrm{MMSE}}$ according to \eqref{eq_MMSE}.
		\STATE Update $\mathbf{p}$ according to \eqref{eq_power_MMSE}.
		\ENDWHILE
		\STATE Initialize $\mathbf{P}^{(0)}=\operatorname{diag}\{\mathbf{p}\}$.
		\STATE Obtain $f(\tilde{\mathbf{u}},\mathbf{P})$ according to \eqref{eq_total_power_MMSE}.
		\FOR   {$t=1:1:\hat{T}_{\max}$}
		\STATE Calculate descent-direction components $\hat{\mathbf{\Psi}}^{(t)}$.
		\FOR   {$m=1:1:\hat{M}$}
		\STATE Calculate candidate descent direction $\hat{\mathbf{d}}^{(t)}_{m}$ in \eqref{eq_position_MMSE_MDD}.
		\STATE Initialize step size $\hat{\tau}^{(t)}_{m}=\bar{\tau}$.
		\STATE Update candidate position vector $\tilde{\mathbf{u}}^{(t)}_{m}$ in \eqref{eq_position_MMSE_candi}.
		\WHILE {\eqref{eq_Arm_MMSE} is not satisfied}
		\STATE Shrink the step size $\hat{\tau}^{(t)}_{m}\leftarrow \hat{\kappa}\hat{\tau}^{(t)}_{m}$.
		\STATE Update candidate position vector $\tilde{\mathbf{u}}^{(t)}_{m}$ in \eqref{eq_position_MMSE_candi}.
		\ENDWHILE
		\ENDFOR 
		\STATE Update MA positioning vector $\tilde{\mathbf{u}}^{(t)}$ according to \eqref{eq_position_MMSE}.
		\STATE Calculate $\mathbf{D}^{(t)}$, $\mathbf{\Psi}^{(t)}$, $\mathbf{b}^{(t)}$, and $\hat{\mathbf{p}}^{(t)}$ according to \eqref{eq_power_MMSE}.
		\STATE Update $\mathbf{P}^{(t)}$ according to \eqref{eq_power_MMSE_it}.
		\IF    {$\left|f\left(\tilde{\mathbf{u}}^{(t)},\mathbf{P}^{(t)}\right)-f\left(\tilde{\mathbf{u}}^{(t-1)},\mathbf{P}^{(t-1)}\right)\right|<\hat{\epsilon}$}
		\STATE Break.
		\ENDIF
		\ENDFOR
		\STATE Set the MA positioning vector as $\tilde{\mathbf{u}}_{\mathrm{MMSE}}=\tilde{\mathbf{u}}^{(t)}$.
		\STATE Set the transmit power as $\mathbf{p}_{\mathrm{MMSE}}=\hat{\mathbf{p}}^{(t)}$.
		\STATE Calculate the MMSE combining matrix $\mathbf{W}_{\mathrm{MMSE}}$ according to \eqref{eq_MMSE}.
		\RETURN $\tilde{\mathbf{u}}_{\mathrm{MMSE}}$, $\mathbf{W}_{\mathrm{MMSE}}$, $\mathbf{p}_{\mathrm{MMSE}}$.
	\end{algorithmic}
\end{algorithm} 

The MMSE-based MDD framework for solving problem \eqref{eq_problem} is summarized in Algorithm \ref{alg_MMSE}, where $\hat{T}_{\max}$ denotes the maximum number of iterations. In line 1, the MA positioning vector is initialized as $\tilde{\mathbf{u}}^{(0)}=\mathbf{0}_{3K}$. Under the given $\tilde{\mathbf{u}}^{(0)}$, the initial $\mathbf{P}^{(0)}$ is obtained by alternately optimizing the receive combining matrix and the transmit power in lines 2-7. Subsequently, the MA positioning vector and the transmit power matrix are alternately optimized in lines 9-26, where the MMSE combining matrix is written as a function of $\tilde{\mathbf{u}}$ and $\mathbf{P}$ for joint optimization. The convergence of Algorithm \ref{alg_MMSE} is analyzed as follows. For each iteration, the update of the MA positioning vector along each candidate descent direction in lines 11-19 can guarantee that the objective value is non-increasing. This is because $f\left(\tilde{\mathbf{u}},\mathbf{P}^{(t-1)}\right)$ and $\rho\left\{\mathbf{D}\mathbf{\Psi}^{-1}\right\}$ are both continuous functions w.r.t. $\tilde{\mathbf{u}}$. If there exists an element in $\hat{\mathbf{d}}^{(t)}_{m}$ which is not equal to zero and the corresponding descent direction is toward the inside of the feasible region, we can always find a sufficiently small positive $\hat{\tau}^{(t)}_{m}$ which can guarantee $f\left(\tilde{\mathbf{u}}^{(t)}_{m},\mathbf{P}^{(t-1)}\right) < f\left(\tilde{\mathbf{u}}^{(t-1)},\mathbf{P}^{(t-1)}\right)$. Besides, since we have $\rho\left\{\mathbf{D}^{(t-1)}(\mathbf{\Psi}^{(t-1)})^{-1}\right\}<1$, a sufficiently small $\hat{\tau}^{(t)}_{m}$ can also guarantee $\rho\left\{\mathbf{D}^{(t)}_{m}(\mathbf{\Psi}^{(t)}_{m})^{-1}\right\}<1$ due to the continuity of function $\rho\left\{\mathbf{D}\mathbf{\Psi}^{-1}\right\}$. Thus, during the iterations, we can always find a new solution ensuring
\begin{equation}\label{eq_conver_MMSE1}
	\begin{aligned}
		f\left(\tilde{\mathbf{u}}^{(t)},\mathbf{P}^{(t-1)}\right) \leq f\left(\tilde{\mathbf{u}}^{(t)}_{m},\mathbf{P}^{(t-1)}\right) \leq f\left(\tilde{\mathbf{u}}^{(t-1)},\mathbf{P}^{(t-1)}\right),
	\end{aligned}
\end{equation}
where the second inequality becomes equality at a point with all descent directions equal to zero or a point located on the boundary of the feasible region defined by \eqref{eq_problem_MMSE_b}. Moreover, since $\mathbf{P}^{(t)}$ is updated as $\operatorname{diag}\{\hat{\mathbf{p}}^{(t)}\}$ in line 22, which yields an optimal MMSE combining matrix, $\mathbf{W}_{\mathrm{MMSE}}(\tilde{\mathbf{u}}^{(t)},\mathbf{P}^{(t)})$, for maximizing the achievable rate region of the users under the current transmit power $\hat{\mathbf{p}}^{(t)}$ \cite{TseFundaWC}. This indicates that the SINR of each user achieved by $\mathbf{W}_{\mathrm{MMSE}}(\tilde{\mathbf{u}}^{(t)},\mathbf{P}^{(t)})$ and $\hat{\mathbf{p}}^{(t)}$ is no smaller than that achieved by $\mathbf{W}_{\mathrm{MMSE}}(\tilde{\mathbf{u}}^{(t)},\mathbf{P}^{(t-1)})$ and $\hat{\mathbf{p}}^{(t)}$, i.e., $\gamma_{k}\left(\mathbf{W}_{\mathrm{MMSE}}(\tilde{\mathbf{u}}^{(t)},\mathbf{P}^{(t)}), \hat{\mathbf{p}}^{(t)}\right) \geq \gamma_{k}\left(\mathbf{W}_{\mathrm{MMSE}}(\tilde{\mathbf{u}}^{(t)},\mathbf{P}^{(t-1)}), \hat{\mathbf{p}}^{(t)}\right)$, $1 \leq k \leq K$. In other words, the updated $\mathbf{P}^{(t)}$ provides additional DoFs for decreasing the total transmit power when solving equations $\hat{\gamma}_{k}=\eta_{k}$, $1 \leq k \leq K$. Thus, we have
\begin{equation}\label{eq_conver_MMSE2}
	\begin{aligned}
		f\left(\tilde{\mathbf{u}}^{(t)},\mathbf{P}^{(t)}\right) \leq f\left(\tilde{\mathbf{u}}^{(t)},\mathbf{P}^{(t-1)}\right).
	\end{aligned}
\end{equation}
Combining \eqref{eq_conver_MMSE1} and \eqref{eq_conver_MMSE2}, we know that Algorithm \ref{alg_MMSE} can achieve a non-increasing sequence of the objective value in \eqref{eq_problem_a} during the iterations. Since the total transmit power is lower-bounded by zero, we can conclude that the convergence of Algorithm \ref{alg_MMSE} for solving problem \eqref{eq_problem} is guaranteed.

In Algorithm \ref{alg_MMSE}, the main computational complexity is caused by the iterations in lines 9-26. Specifically, the calculation of all descent-direction components based on first-order methods in line 10 entails a maximum computational complexity of $\mathcal{O}(KN^{3}\hat{Q})$. Denoting the maximum number of iterations for backtracking line search in lines 15-18 as $\hat{I}_{\max}$, the total computational complexity for searching all candidate descent directions is given by $\mathcal{O}(\hat{I}_{\max}N^{3}\hat{M})$. As a result, the maximum computational complexity of Algorithm \ref{alg_MMSE} for solving problem \eqref{eq_problem} is $\mathcal{O}\left(\hat{T}_{\max}(KN^{3}\hat{Q}+\hat{I}_{\max}N^{3}\hat{M})\right)$.

\subsection{Alternative Solution for the Single-User Case}
In this subsection, we consider a special case where only a single user with MA is served by the BS. As such, the ZF and MMSE combining methods are both degraded into the maximum ratio combining (MRC) because no interference exists. Thus, the optimal MRC vector of the BS is given by
\begin{equation}\label{eq_MRC}
	\mathbf{w}_{\mathrm{MRC}}(\mathbf{u})= \mathbf{h}(\mathbf{u}),
\end{equation}
where $\mathbf{h}(\mathbf{u}) \in \mathbb{C}^{N \times 1}$ denotes the channel vector between the BS and the user defined in \eqref{eq_channel}, and $\mathbf{u}=[x,y,z]^{\mathrm{T}}$ is the MA position of the user. The receive SNR of the signal can be obtained as
\begin{equation}\label{eq_SNR_MRC}
	\tilde{\gamma}_{k} = \frac{\left|\mathbf{w}_{\mathrm{MRC}}(\mathbf{u})^{\mathrm{H}}\mathbf{h}(\mathbf{u})\right|^{2} p}{\left\|\mathbf{w}_{\mathrm{MRC}}(\mathbf{u})\right\|_{2}^{2} \sigma^{2}} = \frac{\left\|\mathbf{h}(\mathbf{u})\right\|_{2}^{2} p}{\sigma^{2}},
\end{equation}
where $p$ is the transmit power of the user. To satisfy constraint \eqref{eq_problem_b}, the minimum transmit power of the user should be given by
\begin{equation}\label{eq_power_MRC}
	\tilde{p}=\frac{\eta \sigma^{2}}{\left\|\mathbf{h}(\mathbf{u})\right\|_{2}^{2}},
\end{equation}
where $\eta=2^{r}-1$ and $r$ represents the minimum-achievable-rate requirement for the user. To minimize the transmit power of the user, we only need to maximize the gain of the channel vector, i.e.,
\begin{subequations}\label{eq_problem_MRC}
	\begin{align}
		\mathop{\max}\limits_{\mathbf{u}}~~~
		&\left\|\mathbf{h}(\mathbf{u})\right\|_{2}^{2} \label{eq_problem_MRC_a}\\
		\mathrm{s.t.}~~~~ &\mathbf{u} \in \mathcal{C}, \label{eq_problem_MRC_b}
	\end{align}
\end{subequations}
where $\mathcal{C}=\mathcal{C}_{1}$ is the region for moving the antenna of the user, i.e., user 1. Problem \eqref{eq_problem_MRC} can be solved by using the proposed MDD framework similar to the procedures in Sections III-A and III-B.

\section{Simulation Results}
In this section, we show simulation results to evaluate the performance of the proposed MA-enabled multiple access systems. We first provide the simulation setup and benchmark schemes, and then present the numerical results for verifying the efficacy of the proposed algorithms.

\begin{table*}[t]\small
	\caption{Simulation Parameters}\label{tab:para}
	\begin{center}
		\begin{tabular}{|c|c|c|}
			\hline
			Parameter            & Description                                                                                  & Value        \\ \hline
			$N_{1} \times N_{2}$ & Antenna array size at the BS                                                                 & 4 $\times$ 4 \\ \hline
			$K$                  & Number of users                                                                              & 12           \\ \hline
			$L$                  & Number of channel paths for each user                                                        & 6            \\ \hline
			$S$                  & Total number of AoAs at the BS                                                               & 20           \\ \hline
			$\lambda$            & Carrier wavelength                                                                           & 0.01 m      \\ \hline
			$g_{0}$              & Average channel power gain at reference distance                                             & -40 dB       \\ \hline
			$\alpha$             & Exponent of path loss                                                                        & 2.8          \\ \hline
			$\sigma^{2}$         & Noise power                                                                                  & -80 dBm      \\ \hline
			$J$                  & Length of the sides of moving region $\mathcal{C}_{k}$                                       & 2$\lambda$   \\ \hline
			$r$                  & Minimum-achievable-rate requirement for each user                                            & 3 bps/Hz     \\ \hline
			$\bar{T}_{\max}$     & Maximum number of iterations in Algorithm \ref{alg_ZF}                   				    & 200          \\ \hline
			$\bar{\tau}$         & Initial step size for gradient descent in Algorithm \ref{alg_ZF}       				        & 10           \\ \hline
			$\bar{\kappa}$       & Scaling factor for shrinking the step size in Algorithm \ref{alg_ZF}    				 	    & 0.5          \\ \hline
			$\bar{\xi}$          & Control parameter for backtracking line search in Algorithm \ref{alg_ZF} 			 	    & 0.6          \\ \hline
			$\bar{\epsilon}$     & Decrement threshold for terminating Algorithm \ref{alg_ZF}                					& 10$^{-6}$    \\ \hline
			$\hat{T}_{\max}$     & Maximum number of iterations in Algorithm \ref{alg_MMSE}                  				    & 200          \\ \hline
			$\hat{\tau}$         & Initial step size for gradient descent in Algorithm \ref{alg_MMSE}        				    & 10           \\ \hline
			$\hat{\kappa}$       & Scaling factor for shrinking the step size in Algorithm \ref{alg_MMSE}      				    & 0.5          \\ \hline
			$\hat{\xi}$          & Control parameter for backtracking line search in Algorithm \ref{alg_MMSE} 			   		& 0.6          \\ \hline
			$\hat{\epsilon}$     & Decrement threshold for terminating Algorithm \ref{alg_MMSE}              				 	& 10$^{-6}$    \\ \hline
		\end{tabular}
	\end{center}
\end{table*}

\subsection{Simulation Setup and Benchmark Schemes}
In the simulation, the users are uniformly distributed around the BS with the distance randomly generated from $d_{\min}=20$ to $d_{\max}=100$ meters (m), i.e., $d_{k}^{2} \sim \mathcal{U} [d_{\min}^{2}, d_{\max}^{2}]$, $1 \leq k \leq K$. We adopt the channel model in \eqref{eq_channel}, where the numbers of transmit and receive channel paths for each user are the same, i.e., $L_{k}^{\mathrm{t}}=L_{k}^{\mathrm{r}}=L$, $1 \leq k \leq K$. For each user, the PRM, $\mathbf{\Sigma}_{k}=\operatorname{diag}\{\sigma_{1}, \sigma_{2}, \cdots, \sigma_{L}\}$, is a diagonal matrix with each diagonal element following CSCG distribution $\mathcal{CN}(0, c_{k}^{2}/L)$, where $c_{k}^{2}=g_{0}d_{k}^{-\alpha}$ is the expected channel power gain of user $k$, $g_{0}$ denotes the expected value of the average channel power gain at the reference distance of 1 m, and $\alpha$ represents the path-loss exponent. Note that for a fair comparison, the total power of the elements in the PRM is the same for the channels of a user with different numbers of paths, i.e., $\mathbb{E}\{\operatorname{tr}(\mathbf{\Sigma}_{k}^{\mathrm{H}}\mathbf{\Sigma}_{k})\} \equiv c_{k}^{2}$. The elevation and azimuth AoDs of the channel paths for each user are random variables with the joint probability density function (PDF) $f_{\mathrm{AoD}}(\theta_{k}^{\mathrm{t}}, \phi_{k}^{\mathrm{t}})=\frac{\cos \theta_{k}^{\mathrm{t}}}{2\pi}$, $\theta_{k}^{\mathrm{t}} \in [-\pi/2, \pi/2]$, $\phi^{\mathrm{t}} \in [-\pi/2, \pi/2]$, which indicates that the AoD has the same probability for all directions in the front half-space of antenna \cite{zhu2022MAmodel}. Due to limited scatters around the BS, the AoAs of channel paths for multiple users are randomly selected from the same set of angles. Specifically, for each channel realization, $S$ pairs of elevation and azimuth AoAs are randomly generated with the joint PDF $f_{\mathrm{AoA}}(\theta^{\mathrm{r}}, \phi^{\mathrm{r}})=\frac{\cos \theta^{\mathrm{r}}}{2\pi}$, $\theta^{\mathrm{r}} \in [-\pi/2, \pi/2]$, $\phi^{\mathrm{r}} \in [-\pi/2, \pi/2]$. Then, the $L$ elevation and azimuth AoAs for each user are randomly selected out of the $S$ candidates. The spatial region for moving the antenna at each user is set as a cubical area of size $[-J/2,J/2] \times [-J/2,J/2] \times [-J/2,J/2]$. The minimum-achievable-rate requirements for all $K$ users are set as the same value, i.e., $r_{k}=r$, $1 \leq k \leq K$, in bits per second per hertz (bps/Hz). The adopted settings of simulation parameters are provided in Table \ref{tab:para}, unless specified otherwise. Each point in the simulation figures is the average result over $10^3$ user distributions and channel realizations.

In this section, the solutions obtained by Algorithms \ref{alg_ZF} and \ref{alg_MMSE} are termed as ``Proposed MA-ZF'' and ``Proposed MA-MMSE'', respectively. For the MDD framework, we set $\bar{M}=\hat{M}=3$ candidate descent directions for both Algorithms \ref{alg_ZF} and \ref{alg_MMSE} in the simulation. Specifically, the opposite direction of gradient, the opposite gradient with momentum, and the descent direction of the quasi-Newton's method are employed as $\bar{Q}=\hat{Q}=3$ descent-direction components. Take the calculation of descent-direction components for the ZF-based MDD framework as example, while that for the MMSE-based MDD framework can be calculated similarly. The gradient of function $g\left(\tilde{\mathbf{u}}\right)$ at location $\tilde{\mathbf{u}}^{(t)}$ can be calculated according to the definition, i.e.,
\begin{equation}\label{eq_grad_ZF}
	\begin{aligned}
	&\left[\nabla_{\tilde{\mathbf{u}}} g\left(\tilde{\mathbf{u}}^{(t)}\right)\right]_{i}=\frac{\partial g\left(\tilde{\mathbf{u}}\right)}{\partial \left[\tilde{\mathbf{u}}\right]_{i}}{\bigg|}_{\tilde{\mathbf{u}}=\tilde{\mathbf{u}}^{(t)}}\\
	&=\lim \limits_{\delta\rightarrow 0} \frac{g\left(\tilde{\mathbf{u}}^{(t)}+\delta \mathbf{e}_{3K}^{i}\right)-g\left(\tilde{\mathbf{u}}^{(t)}\right)}{\delta},~1 \leq i \leq 3K,
	\end{aligned}
\end{equation}
where $\mathbf{e}_{3K}^{i}$ is a $3K$-dimensional vector with a one as the $i$-th element and zeros elsewhere. The gradient with momentum is defined as the weighted sum of gradients during the iterations, i.e.,
\begin{equation}\label{eq_grad_ZF_momentum}
	\mathbf{q}^{(t)} = \beta\mathbf{q}^{(t-1)}+(1-\beta)\nabla_{\tilde{\mathbf{u}}} g\left(\tilde{\mathbf{u}}^{(t)}\right),
\end{equation}
where $\beta$ is a constant weight which is set as $\beta=0.9$. The momentum is initialized as $\mathbf{q}^{(0)}=\mathbf{0}_{3K}$. The descent direction of the quasi-Newton's method is obtained by adopting the widely used Broyden–Fletcher–Goldfarb–Shanno (BFGS) method to approximate the inverse of the Hessian matrix during the iterations, i.e., 
\begin{equation}\label{eq_grad_ZF_Newton}
	\mathbf{a}^{(t)} = -\mathbf{B}^{(t)}\nabla_{\tilde{\mathbf{u}}} g\left(\tilde{\mathbf{u}}^{(t)}\right),
\end{equation}
with {\small$\mathbf{B}^{(t)} = \left(\mathbf{I}_{3K}- \frac{\mathbf{x}_{t} \mathbf{y}_{t}^{\mathrm{T}}} {\mathbf{y}_{t}^{\mathrm{T}} \mathbf{x}_{t}} \right)
\mathbf{B}^{(t-1)} 
\left(\mathbf{I}_{3K}- \frac{\mathbf{y}_{t} \mathbf{x}_{t}^{\mathrm{T}}} {\mathbf{y}_{t}^{\mathrm{T}} \mathbf{x}_{t}} \right) + \frac{\mathbf{x}_{t}\mathbf{x}_{t}^{\mathrm{T}}}{\mathbf{y}_{t}^{\mathrm{T}}\mathbf{x}_{t}}$}, $\mathbf{x}_{t}=\tilde{\mathbf{u}}^{(t)}-\tilde{\mathbf{u}}^{(t-1)}$, $\mathbf{y}_{t}=\nabla_{\tilde{\mathbf{u}}} g\left(\tilde{\mathbf{u}}^{(t)}\right) - \nabla_{\tilde{\mathbf{u}}} g\left(\tilde{\mathbf{u}}^{(t-1)}\right)$ for $t \geq 2$ and $\mathbf{B}^{(1)}=\mathbf{I}_{3K}$. As such, $\bar{\mathbf{\Psi}}^{(t)}$ in line 10 of Algorithm \ref{alg_ZF} is obtained as $\bar{\mathbf{\Psi}}^{(t)}=[-\nabla_{\tilde{\mathbf{u}}} g\left(\tilde{\mathbf{u}}^{(t)}\right), -\mathbf{q}^{(t)}, \mathbf{a}^{(t)}]$.
Moreover, combination weight vectors in \eqref{eq_position_ZF_MDD} and \eqref{eq_position_MMSE_MDD} are set as $\bar{\mathbf{c}}_{1}=\hat{\mathbf{c}}_{1}=[1,0,0]^{\mathrm{T}}$, $\bar{\mathbf{c}}_{2}=\hat{\mathbf{c}}_{2}=[0,1,0]^{\mathrm{T}}$, and $\bar{\mathbf{c}}_{3}=\hat{\mathbf{c}}_{3}=[0,0,1]^{\mathrm{T}}$, respectively\footnote{The descent-direction components themselves are set as candidate descent directions because they imply the significant information of directions for decreasing the objective function. Other values of the combination weight vectors can also be adopted to expand the set of candidate descent directions, which may help obtain better solutions for the MA positioning vector, at the sacrifice of higher computational complexities.}.

Two benchmark schemes for antenna position optimization are defined as follows. For the AS scheme, it is assumed that four FPAs are employed at each user with half-wavelength spacing, where the greedy search \cite{zhong2009AntennaS} is utilized for selecting the best antenna element of each user sequentially for minimizing the total transmit power. For the maximum channel power (MCP) scheme, the MA of each user is deployed at the position which maximizes its channel power, i.e., $\mathbf{u}_{k}^{\star}=\mathop{\arg \max} \limits_{\mathbf{u}_{k} \in \mathcal{C}_{k}} \|\mathbf{h}_{k}(\mathbf{u}_{k})\|_{2}^{2}$, $1 \leq k \leq K$. For both two schemes, the ZF and MMSE combining methods are employed at the BS for calculating the minimum transmit power of users. The ``AS-ZF'' and ``AS-MMSE'' refer to the AS scheme with ZF and MMSE combining, respectively. The ``MCP-ZF'' and ``MCP-MMSE'' represent the MCP scheme with ZF and MMSE combining, respectively. Moreover, the ``MA-MMSE without iteration'' scheme employs the solution for MA positioning vector obtained by Algorithm \ref{alg_MMSE}, while the receive combining matrix of the BS is designed by using the MMSE method in lines 2-6 of Algorithm \ref{alg_MMSE} without alternating optimization between $\mathbf{P}$ and $\mathbf{W}_{\mathrm{MMSE}}$.

\subsection{Convergence Evaluation of Proposed Algorithms}
\begin{figure}[t]
	\centering
	\includegraphics[width=\figwidth cm]{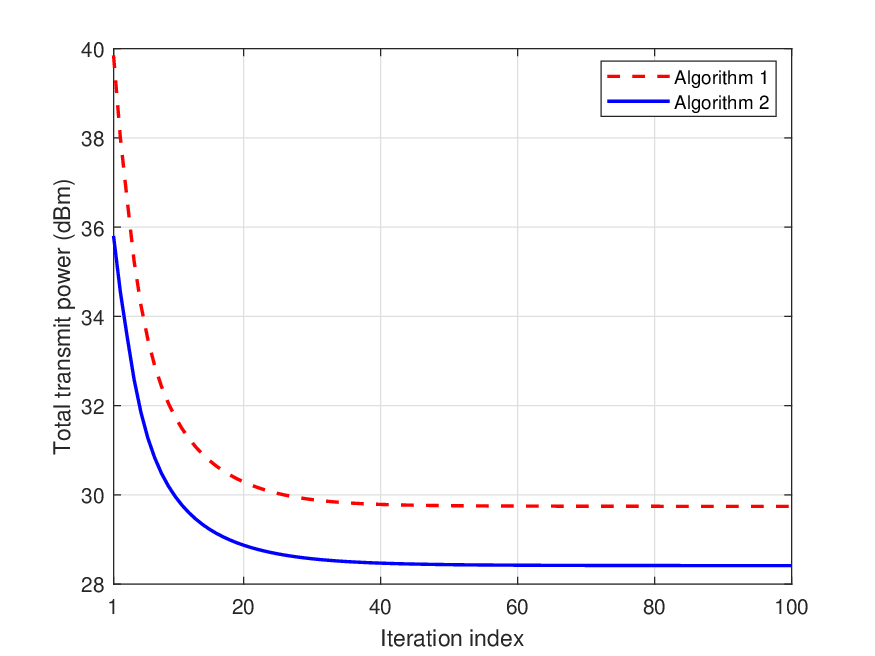}
	\caption{Evaluation of the convergence of the proposed Algorithms \ref{alg_ZF} and \ref{alg_MMSE}.}
	\label{fig:Ita}
\end{figure}

First, in Fig. \ref{fig:Ita}, we evaluate the convergence of the proposed algorithms for MA-enabled multiple access systems. As can be observed, the total transmit powers obtained by Algorithms \ref{alg_ZF} and \ref{alg_MMSE} both decrease with the iteration index and the curves reach a steady state after 40 iterations. The results validate the convergence analysis in Sections III-A and III-B. For the ZF-based solution in Algorithm \ref{alg_ZF}, the total transmit power of users decreases from 39.8 dBm to 29.7 dBm, which yields about 90\% power-saving. For the MMSE-based solution in Algorithm \ref{alg_MMSE}, the total transmit power of users decreases from 35.8 dBm to 28.4 dBm, which yields about 81\% power-saving. Besides, we can find that Algorithm \ref{alg_MMSE} outperforms Algorithm \ref{alg_ZF} in terms of minimizing the total transmit power of users. This is because the MMSE combining can achieve a good balance between the interference and noise power, while the ZF combining results in noise amplification by completely canceling the multiuser interference.

\begin{figure}[t]
	\centering
	\includegraphics[width=\figwidth cm]{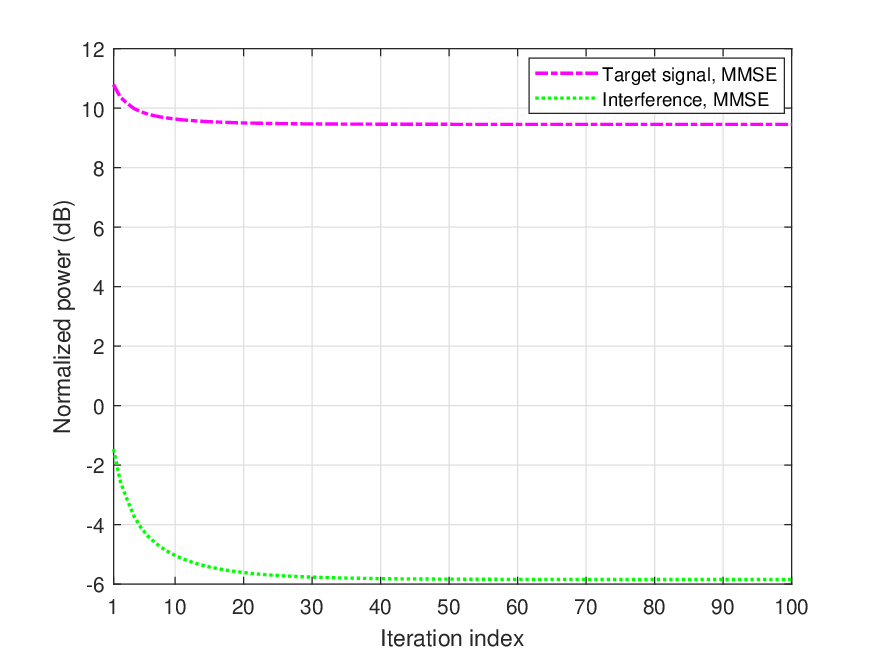}
	\caption{Average receive powers of the target signals and interference for multiple users normalized by noise power versus iteration index.}
	\label{fig:Ita_power}
\end{figure}

To shed more light on the properties of Algorithm \ref{alg_MMSE}, in Fig. \ref{fig:Ita_power}, we show the change of normalized receive powers of the target signals and interference during the iterations, where the powers of the target signal and interference for each user are normalized by noise power, i.e., the average values of $A_{k,k}p_{k}/b_{k}$ and $\sum_{q=1, q \neq k}^{K} A_{k,q} p_{q}/b_{k}$ in \eqref {eq_SINR_MMSE}. During the iterations, the normalized signal power decreases by 1 dB, which is mainly caused by the decreasing transmit power of the users. In comparison, the normalized interference power is reduced by more than 4 dB. The results indicate that the positioning optimization for MAs of the users can effectively decrease the correlation of the channel vectors, which contributes to more effectively mitigating the interference among multiple users. It is foreseeable that if sufficient DoFs are available for moving antennas such that the channel vectors of users are orthogonal to each other, the MRC will become the optimal combining at the BS for each user to maximize the effective channel gain without inducing any interference. In practical communication systems with limited space/DoFs, the MAs of users can still be moved within a local region for reshaping the MAC matrix and improving the communication performance.

\subsection{Channel Characteristics under Antenna Position Optimization}
\begin{figure}
	\centering
	\includegraphics[width=\figwidth cm]{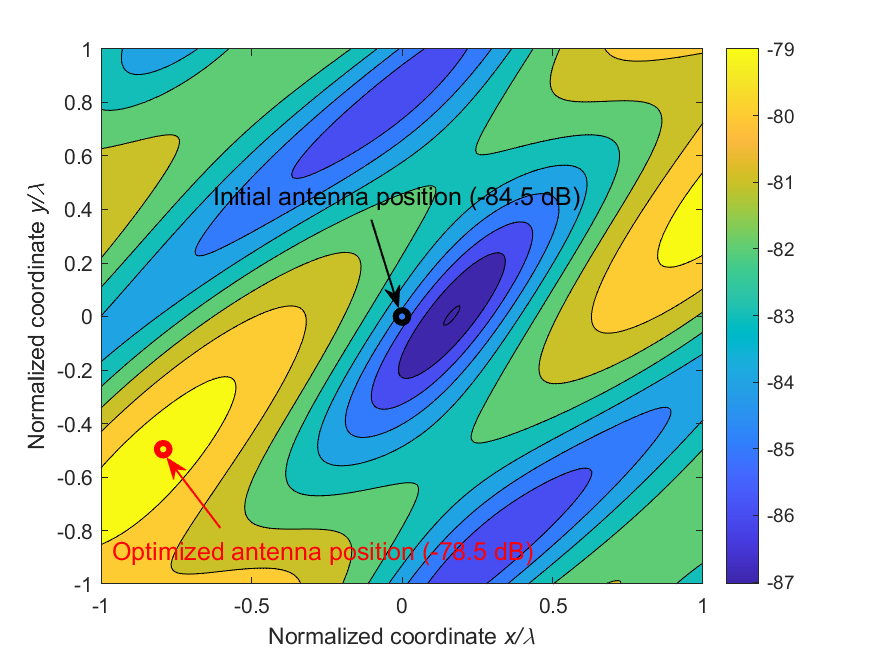}
	\caption{Channel power gain (dB) between the antenna array of the BS and the MA of a single user.}
	\label{fig:ChannelPattern}
\end{figure}

To further explore the impact of MA positioning vector on improving the channel conditions, we show in Fig. \ref{fig:ChannelPattern} one realization of the channel power gain (in dB) versus the MA's position for the single-user case. For ease of presentation, the transmitter region for moving antenna at the user is set as a 2D square, $[-\lambda, \lambda] \times [-\lambda, \lambda]$. As can be observed, due to the prominent small-scale fading in the spatial domain, a small distance of antenna movement (in the order of sub-wavelength) may lead to a great change of the channel response. Specifically, the position of MA is initialized at the origin with channel power gain of $-84.5$ dB. By employing the proposed algorithms, the optimized position of MA is $[-0.8\lambda, -0.5\lambda]$, which yields a local maximum of the channel power gain and achieves a power increase of 6 dB compared to that of the initial position. The results validate the solution for the single-user case in Section III-C, where the minimization of the transmit power of the user is equivalent to maximizing its channel power gain with the BS.


\begin{figure}[t]
	\centering
	\begin{minipage}[t]{1\linewidth}
		\centering
		\includegraphics[width=7.0 cm]{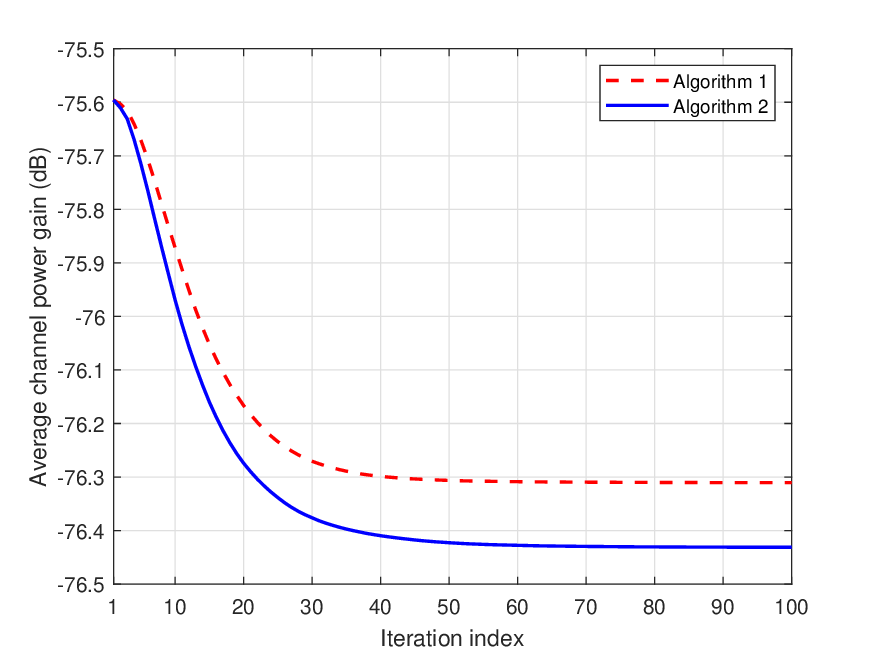}\\
		\small{(a)}
	\end{minipage}\\
	\begin{minipage}[t]{1\linewidth}
		\centering
		\includegraphics[width=7.0 cm]{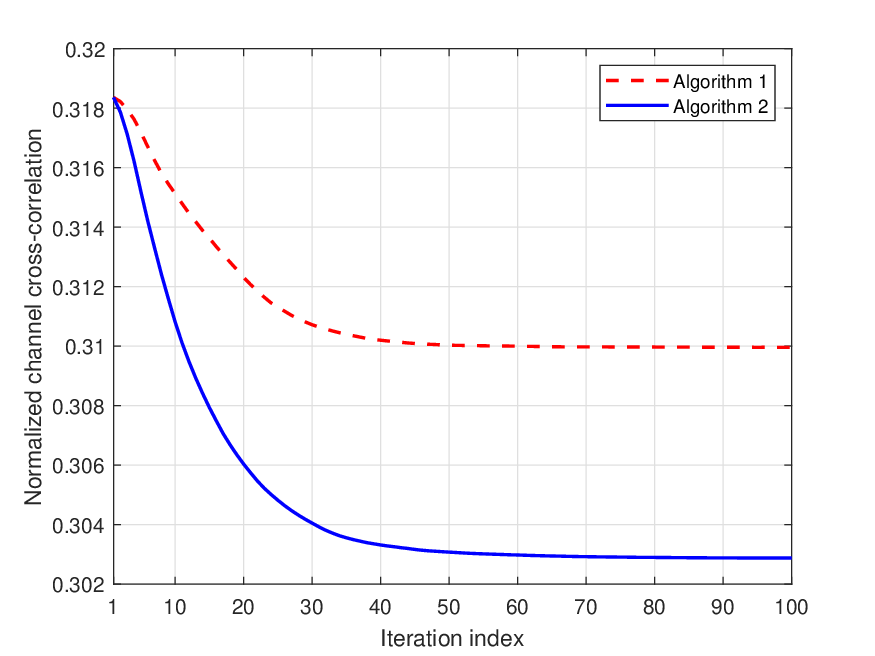}\\
		\small{(b)}
	\end{minipage}
	\caption{Channel characteristics under the position optimization of MAs for multiple users: (a)Average channel power gain; (b) Normalized channel cross-correlation.}
	\label{fig:ChannelPower&ChannelCorr}
\end{figure}

For the multiple-user case, we evaluate in Fig. \ref{fig:ChannelPower&ChannelCorr} the average channel power gain and the normalized channel cross-correlation under antenna position optimization, where the number of users is set as $K=12$. The average channel power gain and the normalized channel cross-correlation are defined as the expected values of $\frac{1}{K} \sum _{1 \leq k \leq K} \left\| \mathbf{h}_{k}(\mathbf{u}_{k}) \right\|_{2}^{2}$ and $\frac{1}{K(K-1)} \sum _{1 \leq k \neq q \leq K} \frac{\left| \mathbf{h}_{k}(\mathbf{u}_{k})^{\mathrm{H}} \mathbf{h}_{q}(\mathbf{u}_{q}) \right|}{\|\mathbf{h}_{k}(\mathbf{u}_{k})\|_{2} \|\mathbf{h}_{q}(\mathbf{u}_{q})\|_{2}}$ over random channel realizations, respectively. Different from the single-user case, the average channel power gain of multiple users decreases during the iterations for optimizing MAs' positions, as shown in Fig. \ref{fig:ChannelPower&ChannelCorr} (a). Meanwhile, the normalized channel cross-correlation for both Algorithms \ref{alg_ZF} and \ref{alg_MMSE} decreases with the iteration index, as shown in Fig. \ref{fig:ChannelPower&ChannelCorr} (b). The above results indicate that the joint position optimization of MAs for multiple users does not simply maximize the average channel power gain of each user, but also is likely to decrease the correlation of multiple channel vectors such that the multiuser interference can be alleviated. Compared to the ZF-based solution in Algorithm \ref{alg_ZF}, the MMSE-based solution in Algorithm \ref{alg_MMSE} can achieve a better tradeoff between the average channel power gain and the normalized channel cross-correlation so as to decrease the total transmit power of users.

\subsection{Performance Comparison with Benchmark Schemes}
\begin{figure}[t]
		\centering
		\includegraphics[width=\figwidth cm]{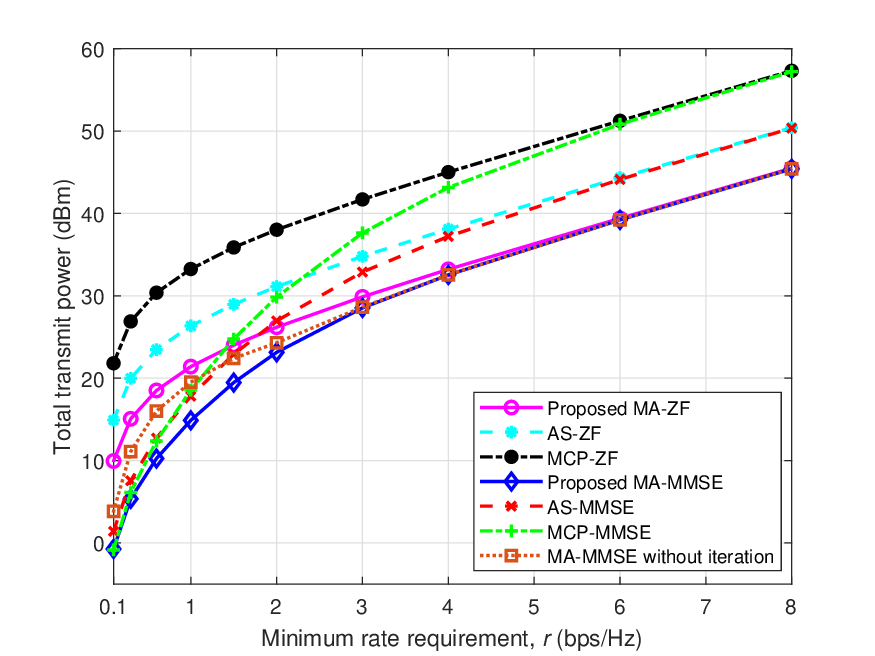}
		\caption{Total transmit powers of different schemes versus minimum-achievable-rate requirements for multiple users.}
		\label{fig:r}
\end{figure}

Next, we compare the performance of the proposed algorithms with benchmark schemes. Fig. \ref{fig:r} shows the total transmit powers of different schemes versus minimum-achievable-rate requirements for the users. We can observe that the proposed MA solutions outperform AS and MCP schemes under both ZF and MMSE combining, especially for a large achievable-rate requirement. If the rate requirement is small, the proposed MA-MMSE solution achieves significantly lower transmit power than the MA-ZF scheme as well as the MA-MMSE scheme without iteration. This is because for small $r$, the SINR of each user is more dominantly affected by the noise. The noise amplification caused by ZF combining may deteriorate the SINR performance compared to MMSE combining. Thus, the receive combining strategy has a dominant influence on the multiuser communication performance in low-SNR region. As the rate requirement increases, the MA-ZF and MA-MMSE (without iteration) schemes can approach the performance achieved by the proposed MA-MMSE in high-SNR region. Besides, we can find that the proposed MA-MMSE solution has a similar performance to the AS-MMSE and MCP-MMSE schemes for small rate requirements because the interference mitigation acquired by MA positioning optimization has a relatively small influence on the SINR performance if the BS receiver operates in low-SNR region. However, as the rate requirement increases to be larger than 2 bps/Hz, the performance gap between the proposed MA-MMSE and AS-MMSE/MCP-MMSE schemes becomes larger.  The results demonstrate that the MA-enhanced multiuser system can obtain higher performance gain than its counterpart of FPA employing AS because of the increasing interference mitigation gain. Moreover, the MCP scheme has the worst performance under both ZF and MMSE combining. This indicates that the performance gain of MA-enabled multiple access systems is mainly achieved by decreasing the multiuser interference rather than increasing the channel power gain of each user.

\begin{figure}[t]
		\centering
		\includegraphics[width=\figwidth cm]{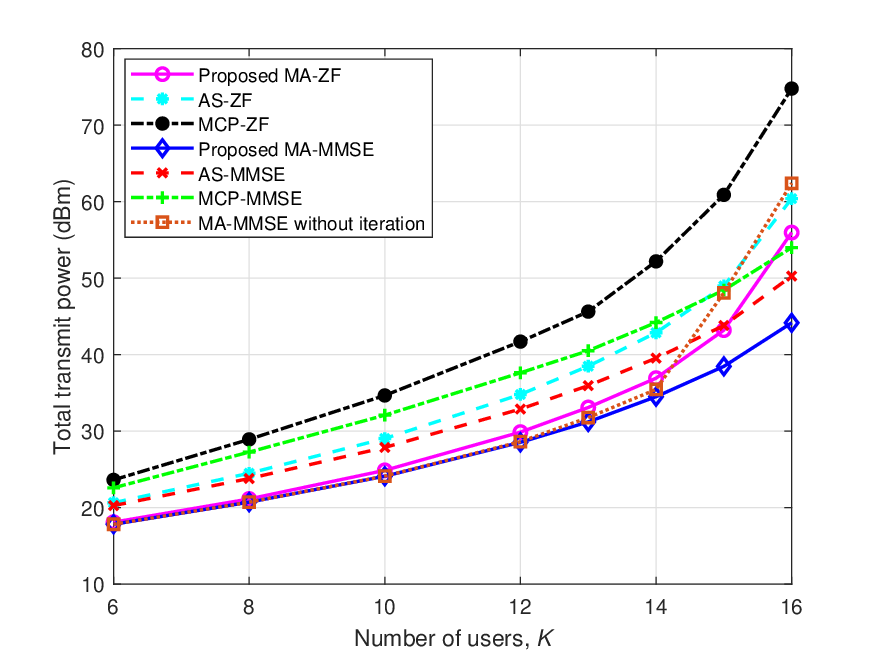}
		\caption{Total transmit powers of different schemes versus the numbers of users.}
		\label{fig:K}
\end{figure}

In Fig. \ref{fig:K}, we compare the total transmit powers of different schemes versus the numbers of users. As can be observed, the total transmit power increases with the number of users for all schemes. The reason is as follows. On one hand, each user needs to send signals to the BS for satisfying its rate requirement, which increases the total transmit power. On the other hand, the increasing number of users results in higher interference for MAC, and thus each user has to increase its transmit power for fulfilling its SINR requirement. If the number of users is small, the multiuser interference at the BS receiver can be satisfactorily suppressed by ZF or MMSE combining, and thus the performance gain provided by MA positioning is not significant. As $K$ increases to be close to the number of antennas at the BS, the interference among multiple users cannot be ideally canceled by receive combining because the correlation of the users' channel vectors is large. In such cases, the positioning optimization of MAs can significantly decrease the correlation of the channel vectors, which is helpful to mitigate multiuser interference. From the results in Fig. \ref{fig:K}, we know that the MA-enabled multiple access systems can obtain higher interference mitigation gain over conventional FPA systems for a large number of users.

\begin{figure}[t]
	\centering
	\includegraphics[width=\figwidth cm]{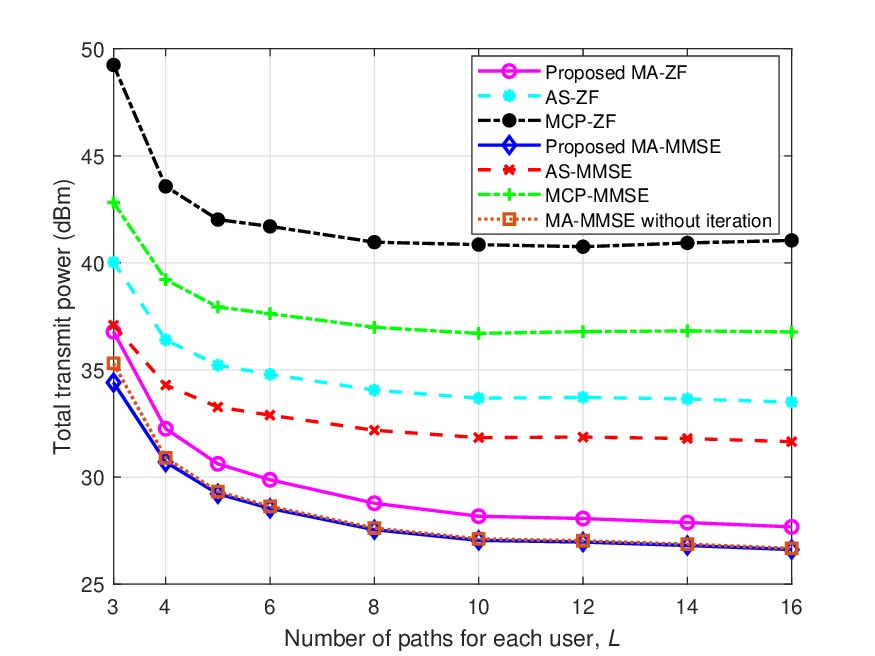}
	\caption{Total transmit powers of different schemes versus the numbers of channel paths for each user.}
	\label{fig:L}
\end{figure}

\begin{figure}[t]
	\centering
	\includegraphics[width=\figwidth cm]{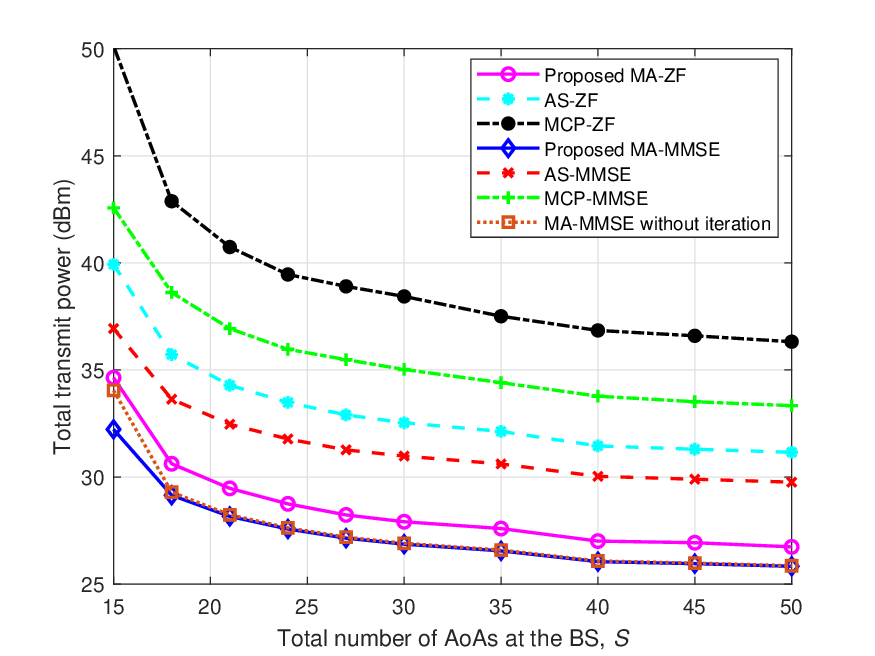}
	\caption{Total transmit powers of different schemes versus the total numbers of AoAs at the BS.}
	\label{fig:S}
\end{figure}

Furthermore, Figs. \ref{fig:L} and \ref{fig:S} show the total transmit powers of different schemes versus the numbers of channel paths for each user and the total numbers of AoAs at the BS, respectively. We can observe again that the total transmit power of MA systems is much smaller than those of AS and MCP schemes because of the interference mitigation gain provided by MA positioning optimization. Meanwhile, the MA-MMSE (without iteration) scheme can achieve a compromised performance between the proposed MA-ZF and MA-MMSE solutions. Besides, the total transmit powers of the proposed and benchmark schemes all decrease with $L$ and $S$. The reason is as follows\footnote{Note that for different values of $L$, the average channel gain of each user with FPA is the same because the power of each channel path is normalized by $L$. Thus, the decrease in total transmit power shown in Fig. \ref{fig:L} is not caused by the increasing average channel gain but due to the decreasing interference.}. As the number of channel paths for each user increases, the spatial diversity is improved in the transmit/receive region, and thus the correlation among the channel vectors for multiple users decreases. Meanwhile, the MA systems can leverage the prominent channel variation to further decrease the correlation of the MAC. However, as shown in Fig. \ref{fig:L}, if $L$ increases to values larger than 10, the descent rate for the total transmit power becomes small. This is because the correlation of channel vectors for multiple users is limited by the number of AoAs at the BS. According to the channel model in \eqref{eq_channel}, the channel vector of user $k$ is a linear combination of the column vectors in receive FRM $\mathbf{F}_{k}$, where the MA positioning optimization can only change the coefficients of linear combinations. If the total number of AoAs of  channel paths for multiple users is limited at the BS side, the FRMs of multiple users have similar column vectors. Thus, the local movement of MAs cannot significantly decrease the MAC correlation. On the contrary, as the total number of AoAs at the BS increases, it becomes easier to decrease the channel correlation among multiple users by moving the antennas.

\begin{figure}[t]
	\centering
	\includegraphics[width=\figwidth cm]{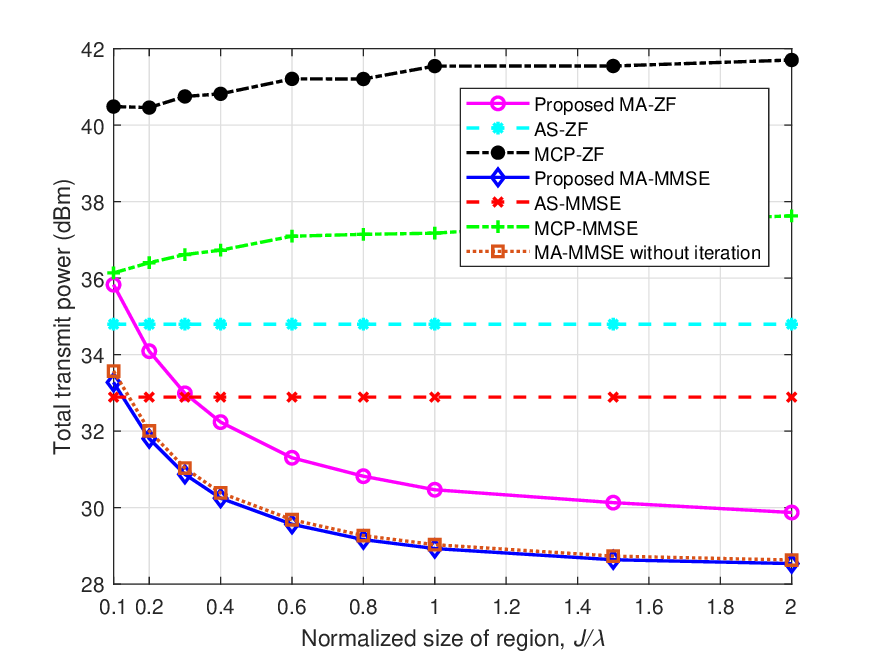}
	\caption{Total transmit powers of different schemes versus normalized region sizes for moving antennas at the users.}
	\label{fig:J}
\end{figure}

In Fig. \ref{fig:J}, we compare the total transmit powers of different schemes versus normalized region sizes for moving antennas at the users, where the size of the moving region is normalized by carrier wavelength, i.e., $J/\lambda$. As can be observed, the total transmit powers achieved by MA systems decrease with the normalized size of the moving region and are much smaller than those of AS and MCP schemes. Moreover, we can see that the total transmit powers of the proposed MA-ZF and MA-MMSE solutions both rapidly decrease as the size of the moving region increases from $0.1\lambda$ to $\lambda$, and achieve their lower bounds for $J=2\lambda$. The results mean that the MA-enabled multiple access systems can even acquire considerable performance gain by moving the antennas of users in a small local region. For example, for communication systems operating at 28 GHz, the volume of a cube of size $2\lambda \times 2\lambda \times 2\lambda$ is about $8 \times 10^{-6}~\text{m}^{3}$, which is implementable in a small device. In addition, we can observe that the MCP schemes achieve higher transmit power as $J$ increases. This indicates that the greedy maximization of the channel power gain for each user may increase the MAC correlation and deteriorate the overall performance.

\subsection{Impact of Imperfect FRI}
The above simulation results are based on the assumption that the BS has a perfect knowledge on the FRI, including the AoDs, AoAs, and PRMs of the channel paths between the BS and users. However, in a real-world communication system, the acquisition of FRI requires to send/receive training pilots and feedback between transceivers. Due to the existence of noise and the limited training overhead, it is challenging to acquire perfect FRI in practice. Thus, it is important to evaluate the impact of imperfect FRI on the performance of MA-enabled multiple access systems. To this end, we introduce two factors to depict the FRI error. On one hand, the AoD error is defined as the difference between the actual and estimated AoDs. Specifically, let $\hat{\vartheta}_{k, j}^{\mathrm{t}}$, $\hat{\varphi}_{k, j}^{\mathrm{t}}$, and $\hat{\omega}_{k, j}^{\mathrm{t}}$ denote the estimated AoDs of the $j$-th channel path between the BS and user $k$, $1 \leq k \leq K$, $1 \leq j \leq L$. The differences between the estimated and actual AoDs are assumed to be random variables following uniform distributions, i.e., $\vartheta_{k, j}^{\mathrm{t}}-\hat{\vartheta}_{k, j}^{\mathrm{t}} \sim \mathcal{U}[-\mu/2, \mu/2$], $\varphi_{k, j}^{\mathrm{t}}-\hat{\varphi}_{k, j}^{\mathrm{t}} \sim \mathcal{U}[-\mu/2, \mu/2$], and $\omega_{k, j}^{\mathrm{t}}-\hat{\omega}_{k, j}^{\mathrm{t}} \sim \mathcal{U}[-\mu/2, \mu/2$], where $\mu$ denotes the maximum AoD error. On the other hand, the PRM error is defined as the difference between the actual and estimated PRMs. Specifically, let $\hat{\mathbf{\Sigma}}_{k}=\operatorname{diag}\{\hat{\sigma_{1}}, \hat{\sigma_{2}}, \cdots, \hat{\sigma_{L}}\}$ denote the estimated PRM between the BS and user $k$, $1 \leq k \leq K$. Considering the estimation error, the difference between the estimated and actual elements in the PRM is assumed to be a CSCG random variable, i.e., $\frac{\sigma_{j}-\hat{\sigma_{j}}}{|\sigma_{j}|} \sim \mathcal{CN}(0,\nu)$, $1 \leq j \leq L$, where $\nu$ is the variance of the normalized PRM error.

\begin{figure}[t]
		\centering
		\includegraphics[width=\figwidth cm]{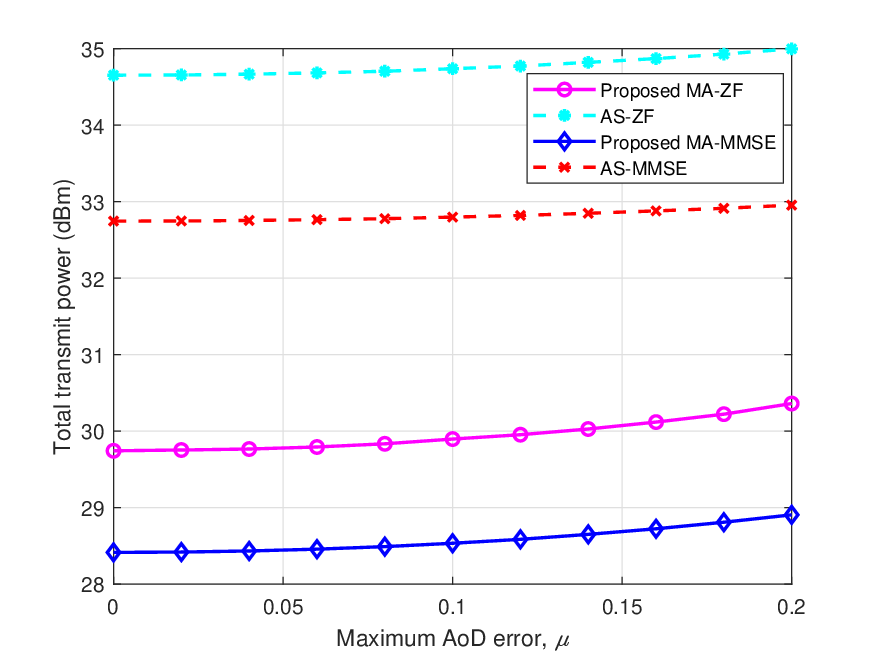}
		\caption{Impact of the AoD error on the performance of the proposed algorithms for MA-enabled multiple access systems.}
		\label{fig:AoD_error}
\end{figure}

In Fig. \ref{fig:AoD_error}, we evaluate the impact of the AoD error on the performance of the proposed algorithms for MA-enabled multiple access systems, where the AoAs of the channel paths and the PRMs between the BS and users are assumed to be known for optimization. In the simulation, the MA positioning optimization is executed based on the estimated (imperfect) FRI. After the MAs are deployed at target positions, the combining matrix at the BS and the transmit power of users are calculated based on the actual channel vectors\footnote{They are estimated separately with given antenna positions of the users and assumed to be perfect in order to focus on investigating the effect of imperfect FRI to the MA position optimization.} between the BS and users. As can be observed in Fig. \ref{fig:AoD_error}, the total transmit power slightly increases with the maximum AoD error because of the deviation between the estimated and actual channels. However, the performance gap between the AS and MA systems is still significant even for large values of $\mu$. Besides, the proposed MA-ZF and MA-MMSE solutions are robust to the AoD error, where the performance loss for $\mu=0.2$ is negligible compared to the perfect FRI case, i.e., not exceeding 0.6 dB power increment. The reason is as follows. Since the size of the region for moving antenna is no larger than $2\lambda$, the deviation of transmit FRVs for estimated and actual AoDs is not significant for a small AoD error. Thus, the MA positioning optimization based on estimated AoDs and transmit FRVs can still yield a good performance for the channels with actual AoDs. 

\begin{figure}[t]
	\centering
	\includegraphics[width=\figwidth cm]{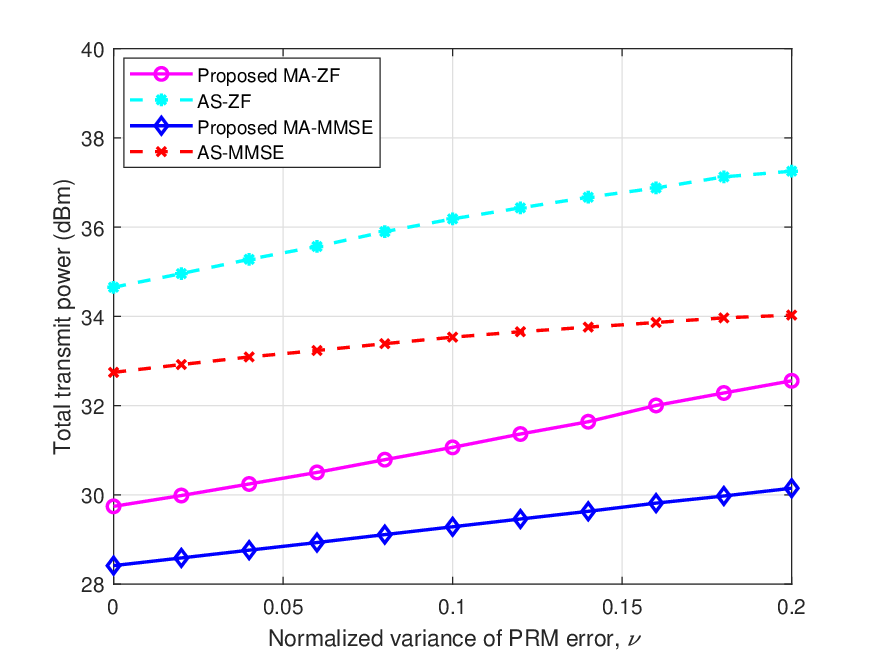}
	\caption{Impact of the PRM error on the performance of the proposed algorithms for MA-enabled multiple access systems.}
	\label{fig:PRM_error}
\end{figure}

Finally, Fig \ref{fig:PRM_error} evaluates the impact of the PRM error on the performance of the proposed algorithms for MA-enabled multiple access systems, where the AoAs and AoDs of the channel paths between the BS and users are assumed to be known for optimization. Similar to Fig. \ref{fig:AoD_error}, in the simulation, the MA positioning optimization is executed based on the estimated FRI, while the combining matrix at the BS and the transmit power of users are calculated based on the actual channel vectors between the BS and users. We can observe again that the total transmit power increases with the variance of normalized PRM error due to the inaccurate channel information for MA positioning optimization. As $\nu$ increases from 0 to 0.2, the total transmit powers of the users for MA-ZF and MA-MMSE solutions are increased by 2.8 dB and 1.7 dB, respectively. However, the superiority of the proposed MA schemes over conventional AS schemes is still significant even for a large value of $\nu$. It is worth noting that the normalized mean square error (MSE) of the estimated PRM is approximately equal to our defined variance of normalized PRM error, i.e., $\|\mathbf{\Sigma}_{k}-\hat{\mathbf{\Sigma}}_{k}\|_{\mathrm{F}}^{2}/\|\mathbf{\Sigma}_{k}\|_{\mathrm{F}}^{2} \approx \nu$. Thus, $\nu=0.2$ indicates a poor condition under which the estimation errors are very close to the values of elements in PRM. In this regard, the proposed MA positioning solutions are shown to be robust to the PRM error. 


\section{Conclusion}
In this paper, we investigated the MA-enhanced MAC for the uplink transmission from multiple users equipped with a single MA to a BS equipped with an FPA array. A field-response based channel model was used to characterize the multi-path channel vector between the antenna array of the BS and each user's MA with a flexible position. Then, we formulated an optimization problem for minimizing the total transmit power of users by jointly optimizing the positions of MAs and the transmit power of users, as well as the receive combining matrix of the BS, subject to a minimum-achievable-rate requirement for each user. Since the resultant problem is highly non-convex and involves coupled variables, we developed two algorithms based on ZF and MMSE combining methods, respectively, where the combining matrix of the BS and the total transmit power of users are expressed as a function of the MA positioning vector during the iterations. For each iteration, the MA positioning vector is updated by using the proposed MDD framework, where multiple candidate descent directions were employed for minimizing the total transmit power and their corresponding step sizes were obtained via backtracking line search. Analytical results showed that the proposed ZF-based and MMSE-based MDD algorithms for MA-enabled multiple access systems can converge to suboptimal solutions with low computational complexities. Besides, an alternative solution for the single-user case was also provided, where the power-minimization problem is equivalent to maximizing the channel gain of the user. Extensive simulations were carried out to verify the superiority of the proposed MA-enabled multiple access systems. It was shown that the proposed MA-ZF and MA-MMSE solutions can significantly decrease the total transmit power of users as compared to conventional FPA systems employing AS. Especially for the scenarios with a large number of users and/or high achievable-rate requirements, the interference mitigation gain provided by MA positioning optimization becomes more significant. Besides, the results revealed that the increasing number of channel paths and the increasing size of the region for moving antennas can help decrease the total transmit power of users. Moreover, we evaluated the impact of imperfect FRI on the solution for MA positioning. It was shown that the proposed algorithms can achieve a robust performance against AoD and PRM errors.

A natural generalization of this work is to consider multi-MA users and/or multi-cell systems. In such scenarios, the coordination of user precoding and its joint optimization with receive combining and MA positioning should be studied. Besides, FRI plays an important role in MA-enabled communication systems, where the AoDs, AoAs, and PRMs should be known to the BS for joint optimization. Thus, efficient FRI estimation protocols and algorithms are required to achieve a good tradeoff between the training overhead and communication performance. These aspects are interesting topics for future research.

\bibliographystyle{IEEEtran} 
\bibliography{IEEEabrv,ref_zhu}

\end{document}